\documentclass{aa}  

\usepackage{graphicx}
\usepackage{txfonts}

\usepackage{natbib}
\bibpunct{(}{)}{;}{a}{}{,} % to follow the A&A style

\begin{document}

   \title{The Wolf-Rayet stars in M31}

   \subtitle{I. Analysis of the late-type WN stars}

   \author{A. Sander
   			  \and
   			  H. Todt
          \and
          R. Hainich
          \and
          W.-R. Hamann
          }

   \institute{Institut f\"ur Physik und Astronomie, Universit\"at Potsdam,
              Karl-Liebknecht-Str. 24/25, D-14476 Potsdam, Germany\\
              \email{ansander@astro.physik.uni-potsdam.de, 
                     htodt@astro.physik.uni-potsdam.de}
             }

	%\date{Received <date> / Accepted <date>}
  \date{Received December 12, 2013; accepted January 22, 2014}

%------- Structured Abstract ----------------------
\abstract{
  Comprehensive studies of Wolf-Rayet stars were performed in the past
  for the Galactic and the LMC population. The results revealed
  significant differences, but also unexpected similarities between the WR populations
  of these different galaxies.
  Analyzing the WR stars in M31 will extend our understanding of 
  these objects in different galactic environments.
}{
  The present study aims at the 
  late-type WN stars in M31. The stellar and wind parameters will tell about the formation
  of WR stars in other galaxies with different metallicity and star formation histories. 
  The obtained parameters will provide constraints to the evolution of massive stars in 
  the environment of M31.
}{
  We used the latest version of the Potsdam Wolf-Rayet model atmosphere code 
  to analyze the stars via fitting optical spectra and photometric data. To account for the
  relatively low temperatures of the late WN10 and WN11 subtypes, our WN models
  have been extended into this temperature regime.
}{
  Stellar and atmospheric parameters are derived for all known late-type WN stars
  in M31 with available spectra. All of these stars still have hydrogen in 
  their outer envelopes, some of them up to 50\% by mass.
  The stars are located on the cool side of the zero age main sequence
  in the Hertzsprung-Russell diagram, while their luminosities range from $10^{5}$ to 
  $10^{6}\,L_{\odot}$. It is remarkable that no star exceeds $10^{6}\,L_{\odot}$.
}{
  If formed via single-star evolution, the late-type WN stars in M31 stem from an initial
  mass range between $20$ and $60\,M_{\odot}$.
  From the very late-type WN9-11 stars, only one star is located in the S Doradus 
  instability strip. We do not find any late-type WN stars with the high 
  luminosities known in the Milky Way.
}


\maketitle

%------------------ Main Paper Content ------------

\section{Introduction}
  \label{sec:intro}
  M31 or the Andromeda galaxy is the largest member of the Local Group,
  which contains only two other spiral-type galaxies, M33 and the Milky Way. 
  Because of the very low foreground extinction 
  towards M31 and the known distance, this galaxy is ideal for studying bright,
  resolvable stellar objects, such as Wolf-Rayet (WR) stars. Analyzing the WR
  stars in M31 will extend our knowledge about the formation and
  evolution of massive stars in other galaxies and allow for a comparative
  analysis of their WR populations. 
  
  Previous studies chiefly focused on the WR population in the Milky Way and the 
  Magellanic Clouds. One of the reasons for this
  limitation was the paucity of available spectra of WR stars in other galaxies.
  The advent of multi-object spectroscopy has greatly facilitated such observations.
  Good quality spectra of extragalactic WR stars are now becoming
  available, allowing for qualitative analyses. A large set of WR spectra for M31
  have been published by \citet{NMG2012} and provided the basis for this work. 
  Photometric data of extragalactic WR stars are contained in the Local Group 
  Galaxy Survey (LGGS) by \citet{Massey+2006}.
  
  Spectroscopically, the WR stars are divided into the WN, WC, and WO subclass. The WN stars
  show prominent nitrogen emission lines in their spectra while the WC and WO stars
  have prominent carbon and oxygen emission lines. In contrast to the WC stars, 
  the oxygen emission lines are significantly stronger in the WO star spectra, including a prominent
  \ion{O}{vi} emission line at $3811$-$34\,$\AA. All subclasses are further split into a sequence 
  of subtypes, defined by the equivalent width or peak ratio of certain emission
  lines \citep{CHS1995, vdH2001}. For the WN stars, these are the subtypes WN2
  to WN11. The WN2 to WN6 subtypes are also called ``early'' types, while the
  WN8 to WN11 are referred to as ``late'' subtypes. The WN7 subtype is something
  in between, although it is often included under the late subtypes. The WN9 to WN11 subtypes
  were introduced by \citet{CHS1995} as a finer replacement of the former Ofpe/WN9 
  classification and are sometimes referred to as ``very late'' WN types (WNVL: very late WN).
  
  In contrast to the early subtypes, late-type WN stars typically have significant 
  amounts of hydrogen \citep{HGL2006}. In the past years, hydrogen-rich WN stars, sometimes classified
  as WNh, have become an interesting topic of research.
  Studies such as \citet{Graefener+2011} have pointed out that at least a significant fraction of
  these stars are not ``classical'' WR stars in the sense that they are stripped cores
  of evolved massive stars. Instead, WN stars of this kind are most likely  
  core-hydrogen burning, and they form the most luminous group within the WR population. The most luminous 
  WR stars in the Large Magellanic Cloud (LMC) and the Milky Way are in fact such hydrogen-rich 
  WN stars \citep{BOH2008,Crowther+2010}.
  
  In this paper we analyze the late-type WN stars with subtypes ranging from WN7 to WN11. 
  Apart from two stars with insufficient spectra,  
  we cover the whole known sample of late-type WN stars in M31 with available optical spectra.
  
\begin{figure}[ht]
  \resizebox{\hsize}{!}{\includegraphics[angle=0]{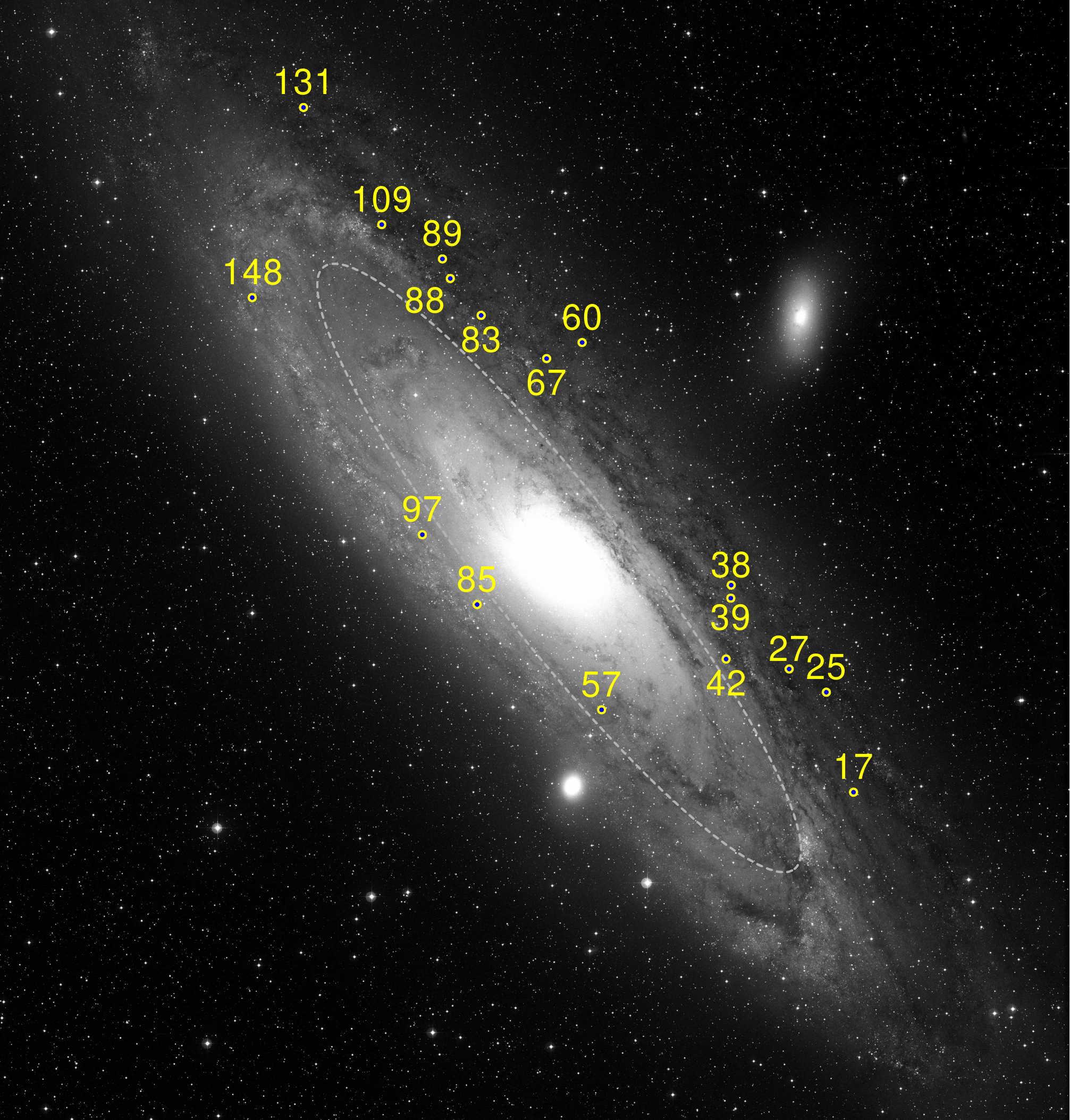}}
  \caption{The positions of the analyzed WN stars in M31. Numbers 
           correspond to the M31WR numbers as in Table\,\ref{tab:fitresults}. 
           The background image is taken from Palomar STScI DSS. The size of the image
           is $\approx 1.9^\circ \times 2^\circ$. The gray dashed ellipse roughly
           corresponds to a radius of $9\,$kpc around the center of M31, corresponding
           to the inner boundary of the region with the most star formation \citep{vdB1964}.}
  \label{fig:wnpos}
\end{figure}

  This present paper builds on our recent work that analyzes the WR star populations in
  the Milky Way \citep{HGL2006,SHT2012} and the LMC \citep{Hainich+2014}.
  Extending the analysis to galaxies beyond the Milky Way may
  in principle allow study of WR stars at higher metallicities. According to \citet{ZKH1994}, 
  the metallicity in M31 has a value of  $\log(\text{O/H}) + 12 = 8.93$, which is a bit higher 
  than the solar value of $8.7$ \citep{EP1995}. However, as in 
  our Milky Way, the metallicity is not uniform across the whole galaxy but instead increases 
  towards the central bulge. Figure\,\ref{fig:wnpos} shows the positions of the analyzed 
  stars in M31. It is significant that none of the known late-type WN stars in M31 are located 
  in the inner part of this galaxy. Instead the stars mostly reside in a zone between 
  9\,kpc and 15\,kpc from the center of M31. It is exactly this region that \citet{vdB1964} found 
  to be the most active in forming stars. Because of their non-central 
  location it might be that the metallicity of our sample is not significantly higher 
  than the one used in our Galactic WN models \citep[cf.][]{HG2004}.
  
  The distance to M31 is well known \citep[$d\,=\,0.77\,$Mpc,][]{CNG}. This offers a big advantage 
  for analyzing of stellar populations. Among the Galactic stars, there is only
  a limited subsample for which distances can be inferred from cluster or association membership.
  Analyzing the M31 WN stars will therefore allow us to crosscheck our findings from the Galactic
  WR analyses and compare them with the LMC results where we have a different type 
  of galaxy and lower metallicity.
  
  In the next section, we briefly characterize the stellar wind models
  used in this work. In Sect.\,\ref{sec:sample} we report the results and compile 
  the obtained stellar parameters. In Sect.\,\ref{sec:evol} we discuss the results and 
  compare them with theoretical stellar evolution 
  tracks. The conclusions are drawn in the Sect.\,\ref{sec:conclusions}. 
  In the appendix (Sect.\,\ref{appsec:specfits}) we show all spectral fits 
  obtained in this work.

%--------- Figure   ----------------------------------------------------
%--------- end Figure ----------------------------------------------------
  
\section{Stellar wind models}
  \label{sec:models}

  We employ stellar atmosphere models computed with the 
  the Potsdam Wolf-Rayet (PoWR) code. For the present study we extend 
  the published WN model grids \citep[e.g.,][]{HG2004}	
  to lower temperatures in order to cover the very late subtypes WN10 and WN11.
  The PoWR models assume a spherically symmetric, stationary
  outflow. The models cover the whole stellar atmosphere and wind, 
  starting at the inner boundary $R_{\ast}$, which is defined at a Rosseland
  optical depth of $\tau = 20$. The stellar temperature $T_{\ast}$ is defined
  via Stefan-Boltzmann's law as the effective temperature at $R_{\ast}$.

  The velocity field is described as a so-called $\beta$-law in the supersonic
  part. Here we use a value of $\beta = 1$. Wind inhomogeneities are treated in the 
  so-called ``microclumping'' approximation assuming optical thin clumps, which have a 
  density value increased by a factor of $D$ compared to a smooth wind with same mass-loss rate. 
  This factor is set to $D = 4$ in all models, similar to our Galactic WN models used in earlier
  works, such as \citet{HGL2006} and \citet{Liermann+2010}. In principle, the PoWR code 
  can account for inhomogeneities of any optical depth \citep{OHF2007}, but we do not
  consider macroclumping in this work. The interclump space is assumed to
  be void, and thus the volume-filling factor $f_{V}$ is simply given by $f_{V} = D^{-1}$.

\begin{table}
  \caption{Fundamental parameters of the WNh model grids}
  \label{tab:grids}
  \centering
  \begin{tabular}{l c c c}
  \hline\hline
    Parameter				  \rule[0mm]{0mm}{3mm}    	  	&	\multicolumn{3}{c}{ WNh grids }  \\  
  \hline
     $X_{\mathrm{H}}$ \rule[0mm]{0mm}{3mm}  &	       50\%  &     35\%    &  20\%   \\
     $X_{\mathrm{He}}$                      &	       48\%  &     63\%    &  78\%   \\
     $X_{\mathrm{C}}$		                    &	\multicolumn{3}{c}{0.01\%}   \\
     $X_{\mathrm{N}}$		                    &	\multicolumn{3}{c}{ 1.5\%}   \\
     $X_{\mathrm{Fe}}$\tablefootmark{a}		  & \multicolumn{3}{c}{0.14\%}   \\[2mm]
     
     $T_{\ast}/\,$kK											  & \multicolumn{3}{c}{ 18..45 }   \\
     $R_{\mathrm{t}}/\,R_{\odot}$           & \multicolumn{3}{c}{ 4..100 }   \\
     $\log L/L_{\odot}$									  	& \multicolumn{3}{c}{ 5.3  }   \\
     $\varv_{\infty}$\,/\,km\,s$^{-1}$      &	\multicolumn{3}{c}{ 500/1000\tablefootmark{b} }  \\
     $D$                                    & \multicolumn{3}{c}{   4  }  \\
  \hline
  \end{tabular}
  \tablefoot{
  	\tablefoottext{a}{Generic element, including the iron group elements Fe, Sc, Ti, V, Cr, Mn, Co, and Ni.
  	                  See \citet{GKH2002} for relative abundances.}
  	\tablefoottext{b}{Standard models in the whole parameter range were available with $500$ and $1000$\,km/s. 
  	                  For certain stars, additional models with especially adjusted (often lower) values 
  	                  were calculated.}
  }  
\end{table}  
  
  As discussed in in \citet{HG2004}, there are two main parameters that describe the
  overall appearance of a normalized Wolf-Rayet emission line spectrum, namely the
  stellar temperature $T_{*}$ and the so-called ``transformed radius''  
  \begin{equation}
  \label{eq:rt}
 R_{\mathrm{t}} = R_{*} \left[ \frac{\varv_{\infty}}{2500\,\mathrm{km/s}} 
 \left/ \frac{\dot{M} \sqrt{D} }{10^{-4} M_{\odot}/\mathrm{yr}} \right. 
 \right]^{\frac{2}{3}},
  \end{equation} 
  which was introduced by \citet{SHW1989} to describe that stars with the same
  value of $R_{\mathrm{t}}$ have roughly the same spectral appearance. While the bulk
  of WN models have been calculated with a terminal wind velocity of $\varv_{\infty} = 1000\,$km/s,
  we adopt a value of $\varv_{\infty} = 500\,$km/s for the latest subtypes (cf. Table\,\ref{tab:grids}). 
  In some cases where the lines clearly indicate even lower values, special models are calculated 
  to further improve the matching of the observed spectral lines.

\section{The studied sample}
  \label{sec:sample}  

\subsection{Name convention}
  \label{subsec:name}

  We analyzed the whole sample of WN stars with late subtypes in M31 stars from 
  \citet{NMG2012}. Their spectra were taken with the 6.5m telescope at the MMT Observatory
  in Arizona using the Hectospec multi-object fiber-fed spectrograph. The useful spectral
  range extends from 3700 to 7500\,\AA. The spectral resolution is $6\,$\AA. 
  Instead of using the long LGGS names from \citet{Massey+2006}, we introduce
  a WR number system, following Table\,5 from \citet{NMG2012}, which has
  already been sorted by coordinates. Thus J003857.19+403132.2 becomes M31WR\,1,
  J003911.04+403817.5 becomes M31WR\,2, and so on. In the text of this work, we further 
  shorten M31WR\,1 to \#1 to improve the readability. The full LGGS names are 
  given in the figure captions and also in the spectral fits in the Online Material.
  
  This study is focused on WN stars that exhibit hydrogen in their wind. This
  comprises all WN8 to WN11 subtypes in the sample from \citet{NMG2012} but also
  the two WN7 stars \#38, \#57, and \#88. The WN7 star \#86 most likely shows a composite
  spectrum (see Sect.\,\ref{subsec:bin}) and is therefore excluded from our analysis.
  For the two other stars classified as WN7, \#1 and \#41, the
  quality of the spectra is not sufficient for a quantitative analysis. The two 
  objects classified as WN6-7, \#10 and \#21, are apparently hydrogen-free and therefore 
  their analyses is postponed to our forthcoming study of the WNE class. 
    
  Our remaining sample thus comprises 17 stars (3 WN7, 1 WN7-8, 6 WN8, 3 WN9, 3 WN10, 1 WN11).  
  Compared to the total number of 92 WN stars in M31 \citep{NMG2012}, the number of WNL stars 
  ($\approx 20$) is surprisingly small. Among the Galactic single WN stars, \citet{HGL2006} found
  that 26 out of 59 objects belong to the WNL class, taking detectable hydrogen as criterion.   
  As M31 is bigger than our own Galaxy, one would expect that the total
  number of WN stars is also higher. We therefore suggest that the known sample of WR stars in M31
  is actually far from being complete, especially for the very late WN9 to WN11 subtypes. A
  lot of these stars have been discovered in the Milky May in the past decade, when infrared 
  spectroscopy became available and obscured clusters could be studied, e.g., in the Galactic
  center \citep{Krabbe+1995,Martins+2008,Liermann+2010} or in the G305 star forming complex \citep{Davies+2012}.
  In the current Galactic WR star catalog\footnote{\texttt{http://pacrowther.staff.shef.ac.uk/WRcat/}},
  there are roughly $60$ WN9 stars listed, but only eight of them have optical photometry, and for
  the rest the classification relies on limited infrared data. One can therefore expect that the 
  current sample of late-type WN stars in M31 is not complete. It is difficult to
  quantify how much we are really missing. If the star formation rate (SFR) was roughly the
  same as in the Milky Way, it would be a large amount. However, if the SFR is significantly lower and
  comparable to M33, as \citet{NMG2012} inferred from the integrated H$\alpha$ luminosity, the 
  number of missing late-type WN stars would propably be only close to our sample size.
  \citet{NMG2012} further state that for WN9-11 stars the problem of missing objects is not only due to
  obscuration, but also to the detection method, which was not sensitive to those subtypes.

\subsection{Binarity}
  \label{subsec:bin}

  It is known that a large fraction of the massive stars are actually in binary systems 
  \citep[see, e.g.,][]{VC1980,Barba+2010,Sana+2013}. The detection of close binarity requires
  detailed monitoring of photometric or radial-velocity variations. Such studies had not yet been
  performed for WR stars in M31. At the large distance of M31, even a 
  small cluster may appear as a single star. Therefore one may ask how many of our sample stars are still
  undetected binaries. In the case of a luminous companion, the spectrum would in fact be composite,
  and the analysis may lead to wrong stellar parameters. 
	
	One indicator of a composite spectrum is the so-called ``dilution'' of WR emission lines. 
	In a system with a WR star and one or more luminous components without significant emission lines, 
	the continua of all stars will add up, while the emission lines originate in the WR component 
	alone. In the normalized spectrum, the emission lines will therefore be significantly weaker 
	compared to a single WR star of the same subtype. 
  While the dilution criterion seems to be a rather good indicator of WC binaries \citep{SHT2012}, 
  \citet{HGL2006} argue that it is not sufficient for WN stars because it 
  is possible to find appropriate stellar atmosphere models for WN stars with ``weaker'' lines.
   
  Even if apparent dilution of emission lines might not be a sufficient binarity criterion,
  the absence of dilution is a rather good indicator that a star at least does not have a luminous
  companion. In our sample of late type WN stars, only two objects show the typical signs of diluted
  emission lines. The WN7 star \#86 shows the typical broad lines with low peak heights, e.g. of \ion{He}{ii}\,4686\,\AA,
  and is therefore excluded here. The situation is less clear for the WN8 star \#109. 
  The relatively broad 
  \ion{N}{iii}\,4634-42\,\AA\ complex cannot be reproduced by the same model that fits the 
  rest of the lines. The luminosity of $\log L/L_{\odot} = 5.9$ is slightly higher than for 
  the majority of stars in our sample. Nevertheless, we feel that these indications are not
  strong enough to exclude \#109 from our sample. 
  
%--------- Figure   ----------------------------------------------------
\begin{figure}[ht]
  \resizebox{\hsize}{!}{
    \includegraphics[angle=0]{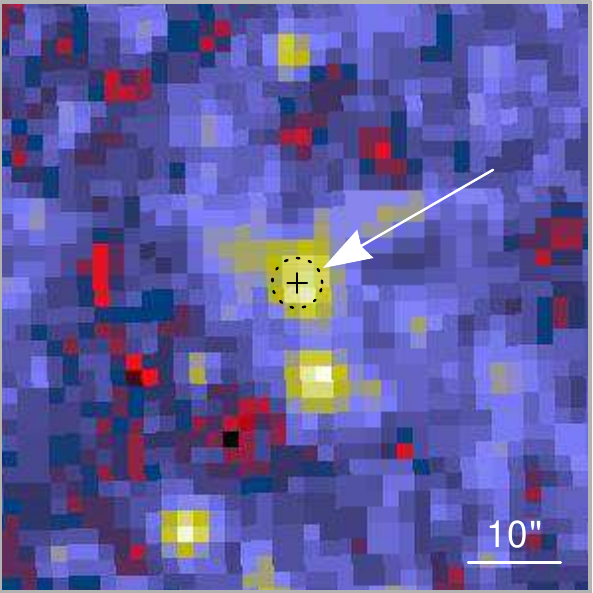}
    \includegraphics[angle=0]{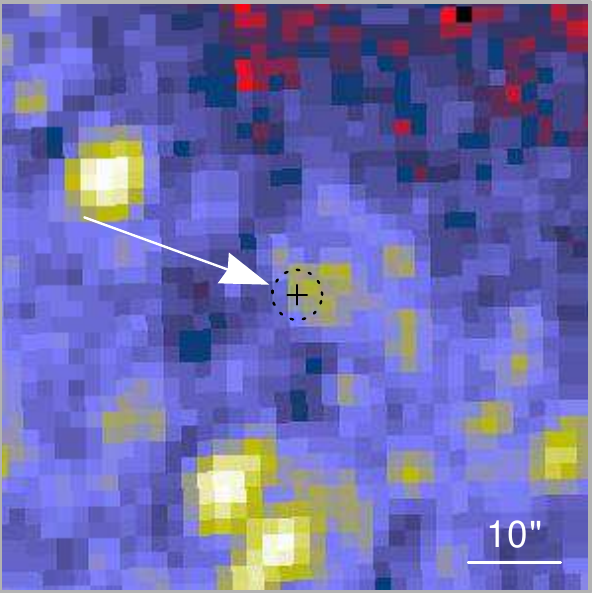}
  }
  \caption{Surroundings of M31WR\,86 (left) and M31WR\,109 (right): The black cross marks the location of the object
           referring to the coordinates in \citet{NMG2012}. The slight offset between the sources in DSS images 
           and \citet{NMG2012} is seen for all WR stars. The dotted 
           circle corresponds to a distance of $10\,$pc around the objects. The images were retrieved with SkyView  
           from the Palomar STScI DSS and have a size of $1.06' \times 1.06'$ each.}
  \label{fig:envbincand}
\end{figure}
%--------- end Figure ----------------------------------------------------
  
  Apart from the spectral appearance, we also took a look at the surroundings of the binary
  candidates using the Palomar STScI DSS images. While other stars in our sample appear as a single
  source, \#86 (left panel in Fig.\,\ref{fig:envbincand}) appears to be in a crowded region
  and might even be a member of a small cluster, so the spectrum is really not from a single Wolf-Rayet
  star, and we exclude the star from further study. The star \#109 (right panel in Fig.\,\ref{fig:envbincand})
  lacks the bright appearance of most other sources, but the situation remains unclear so we keep 
  this star in the sample.

\section{Results}
  \label{sec:results}

\subsection{Fitting the normalized spectra}
  \label{subsec:fit}

  In a first step of our analysis, we identify the best-fitting model for each star 
  from our standard grid in the $T_{*}$-$R_{\mathrm{t}}$-plane. The focus is usually 
  on the unblended helium lines. In several cases, the helium lines are in good agreement 
  with the observation, but the observed hydrogen lines are stronger than predicted by
  our models with $X_\mathrm{H} = 20$\%. For such cases we calculate two additional grids 
  of models with hydrogen abundances of 35\% and 50\%, respectively (see Table\,\ref{tab:grids}).
  Together with our standard WN grid, 
  which uses $X_\mathrm{H} = 20$\%, this allows us to estimate the hydrogen fraction.
  
%--------- Figure   ----------------------------------------------------
\begin{figure*}
  \centering
  \includegraphics[angle=90,width=17cm]{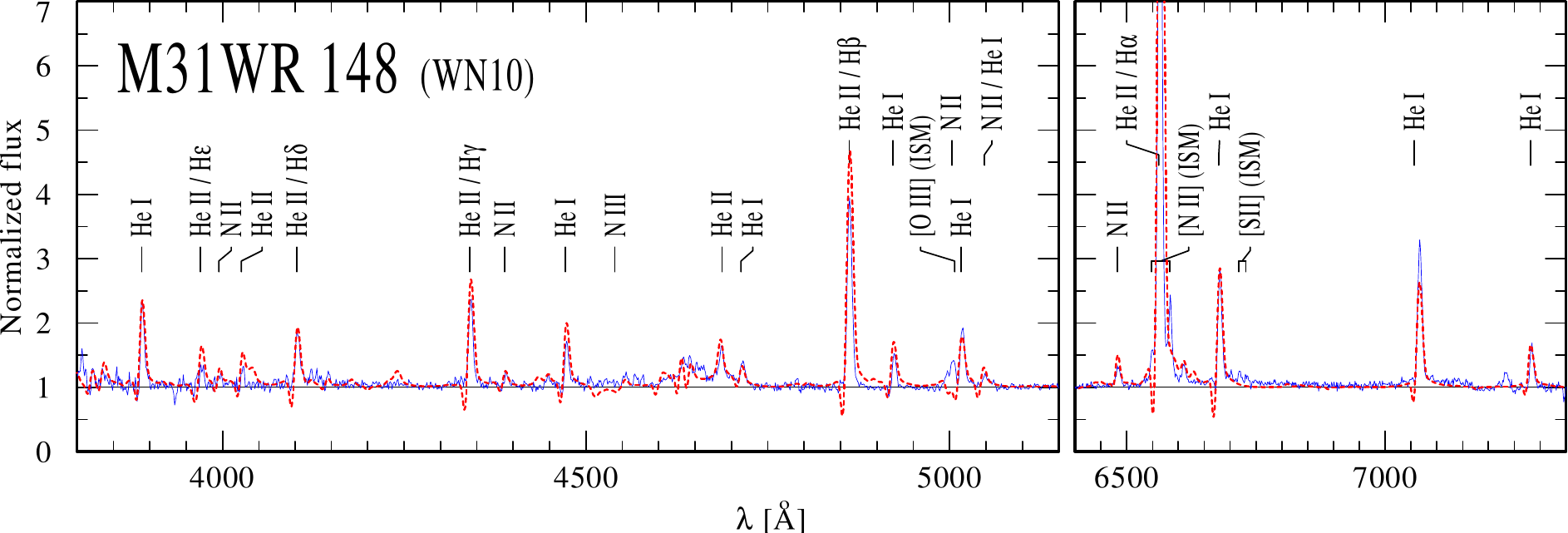}
  \caption{Optical spectrum of M31WR\,148 alias LGGS J004542.26+414510.1. 
    The solid thin blue line shows the observation from \citet{NMG2012}, and the
    best-fitting PoWR model is overplotted as red thick dotted line. Important
    lines, including those from interstellar gas, are identified.}
  \label{fig:specfit}
\end{figure*} 
%--------- end Figure -------------------------------------------------- 

  In a further step, the terminal wind velocity $\varv_{\infty}$ is adjusted if the line width 
  is significantly over- or underestimated by our standard grid models. Individual models are 
  then calculated until the line width is reproduces satisfactorily.
  
  M31 has a mean radial velocity (relative to the sun) of 
  $\varv_{\mathrm{rad}} = -301\,$km/s \citep{CNG}. For our line fits, all observed spectra 
  have to be shifted according to this value. In addition, several stars show a significant
  individual radial velocity. In such cases, we apply an additional shift until the observation 
  matches with the model. The
  resulting peculiar radial velocities are included in Table\,\ref{tab:fitresults}. Obtaining 
  precise radial velocities is not an aim of this study, and therefore we accept a relatively
  large error margin of $\pm 25\,$km/s. Nevertheless, the obtained results are in 
  good agreement with the known rotation of M31 \citep[e.g.,][]{RF1970}.   
  
  Figure \ref{fig:specfit} shows a typical example of a normalized line fit. The example spectrum demonstrates
  how narrow the emission lines of these late-type WN stars are. In this case, a terminal
  velocity of $\varv_{\infty}\,=\,500\,$km/s is used, while other stars have even lower velocities
  (see Table\,\ref{tab:fitresults}).
  
  There are a few lines in the spectra that cannot be
  reproduced by our models. Apart from a terrestrial absorption feature around 5575\,\AA, there are several interstellar 
  emission lines, such as [\ion{O}{ii}]\,3727\,\AA, [\ion{O}{iii}]\,5007\,\AA, [\ion{N}{ii}]\,6548\,\AA\,
  [\ion{N}{ii}]\,6583\,\AA, and [\ion{S}{ii}]\,6717/31\,\AA. These forbidden lines are typical of the so-called
  DIG (diffuse interstellar gas) that is associated with prominent spiral arms in M31 \citep{GWB1997}.
  The two [\ion{N}{ii}] lines at 6548\,\AA\ and 6583\,\AA\ can overlap with H$\alpha$ in early-type WN spectra.
  However, for the late types studied here, H$\alpha$ is narrower and hence well separated from these [\ion{N}{ii}] lines. 

  With the help of the three WNh grids we can estimate the hydrogen content of the sample stars 
  to a precision of approximately $\pm10$\% (cf. Table\,\ref{tab:fitresults}). 
  All stars of our sample contain a detectable amount of hydrogen, mostly with a mass fraction 
  between 20\%\ and 35\%. For four objects, the WN9 star \#27 and the WN8 stars \#39, \#85, and 
  \#109, a model with 50\%\ is required for reproducing the observed strength of the Balmer lines. 
  
%--------- Figure   ----------------------------------------------------
\begin{figure}[ht]
  \resizebox{\hsize}{!}{
    \includegraphics[angle=-90]{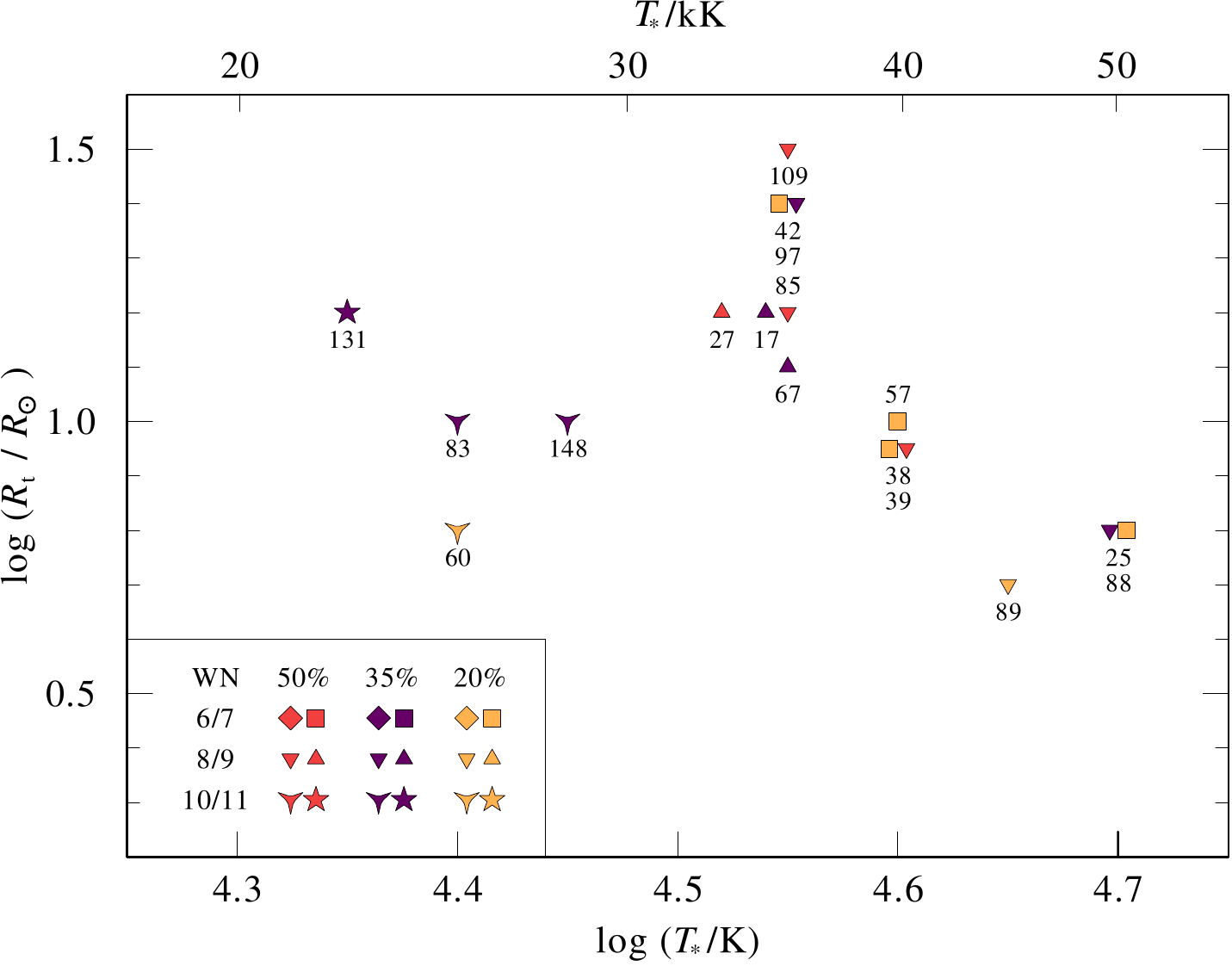}
  }
  \caption{Obtained positions of the analyzed WNh stars in M31 in the $\log\,R_{\mathrm{t}}$-$\log\,T_{\ast}$-plane.
           The different colors indicate the different hydrogen content. Red: $50\%$ hydrogen,
           purple: $35\%$ hydrogen; orange: $20\%$ hydrogen. The type of the symbol reflects the WN subtype with
           a similar meaning as in Fig.\,\ref{fig:mdot-l}. The numbers are the M31WR numbers as in Table\,\ref{tab:fitresults},
           indicating for which star a particular model has been used.}
  \label{fig:wnstats}
\end{figure}
%--------- end Figure ----------------------------------------------------  
  
   From the fit of the normalized spectra we obtain the stellar temperature $T_{\ast}$
   and the transformed radius $R_{\mathrm{t}}$ for all stars in our sample (cf. Table\,\ref{tab:fitresults}
   and Fig.\,\ref{fig:wnstats}). The location of the stars in the $R_{\mathrm{t}}$-$T_{\ast}$-plane
   reflects their spectral subtypes. There is no correlation between the hydrogen content and subtype 
   or temperature.
   
   The terminal wind velocities $\varv_{\infty}$ are higher for stars with less hydrogen. The 
   three WN7 stars \#38, \#57, and \#88 require models with $\varv_{\infty} = 1000\,$km/s in contrast 
   to the rest\footnote{We also give $\varv_{\infty} = 1000\,$km/s for \#109, but 
   as discussed in Sect.\,\ref{subsec:bin}, this object is suspicious and might not 
   be a single star.} of the analyzed WNh stars.
   
   For two stars, \#42 and \#60, the fitting process caused some trouble. The star \#42 has not been classified
   in detail so far because of limited available classification lines. Because of the highly uncertain
   results in this case, this star is excluded from most of the further discussions in this work, and 
   \#60 shows strong \ion{He}{i}-lines that could not be reproduced. Details for both objects are 
   discussed in the Online Appendix (Sect.\,\ref{appsec:objnotes}). The complete set of all spectral 
   fits is located in the following Online Appendix (Sect.\,\ref{appsec:specfits}).   

\subsection{Luminosities}
  \label{subsec:lumo}
  
  After the best-fitting model has been selected by means of the line fit, the luminosity
  of the model now has to be adjusted to reproduce the observed absolute fluxes of the star. The grid 
  models were calculated with a luminosity of $\log L/L_{\odot} = 5.3$ and are now scaled 
  such that the observed spectral energy distribution (SED) is matched. Scaling the luminosity does 
  not affect the fit of the normalized line spectrum, as long as the transformed radius $R_{\mathrm{t}}$
  is kept. 

%--------- Figure   ----------------------------------------------------
\begin{figure*}
  \centering
  \includegraphics[width=17cm]{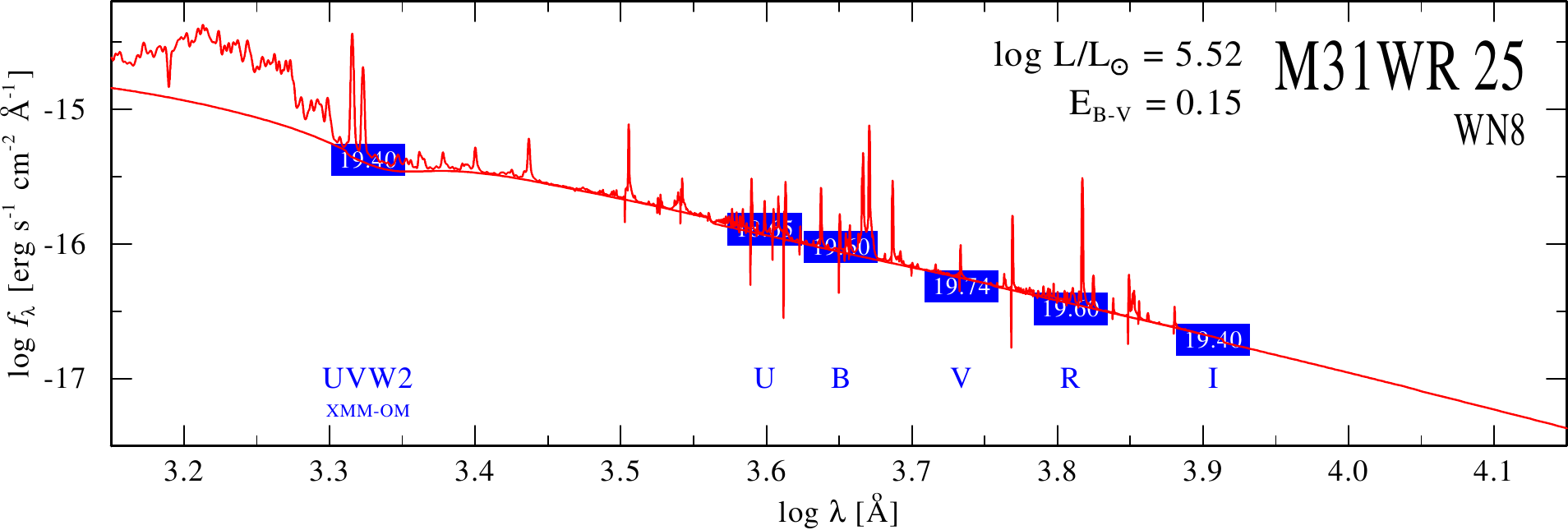}
  \caption{Spectral energy distribution (SED) of M31WR\,25 alias 
    LGGS J004036.76+410104.3 obtained by fitting the observed photometry
    marks (blue boxes) with the PoWR model (red solid line).}
  \label{fig:sedfit}
\end{figure*} 
%--------- end Figure --------------------------------------------------
  
  At the same time, the interstellar reddening color excess $E_{B-V}$ has to be determined, too.
  The reddening is relatively small for all our objects in our sample. We apply the reddening law 
  by \citet{F1999}. For M31 we adopt a distance modulus of D.M.$~= 24.4$
  corresponding to $d = 0.77\,$Mpc \citep{CNG}, and geometrically dilute the model flux accordingly.
    
  We use the UBVRI photometry from \citet{Massey+2006} to describe the spectral energy distribution,
  since the spectra from \citet{NMG2012} are not flux-calibrated. For a few objects
  (\#10, \#21, \#25, and \#42) there are UV flux measurements from the XMM 
  Optical/UV Monitor (XMM-OM) available from \citet{Page+2012}. An example of an
  SED fit is shown in Fig.\,\ref{fig:sedfit}, the resulting luminosities and $E_{B-V}$ color
  excesses are compiled in Table\,\ref{tab:fitresults}.

\subsection{Mass-loss rates}
  \label{subsec:mdot}	 
 
\begin{figure}[ht]
  \resizebox{\hsize}{!}{\includegraphics[angle=0]{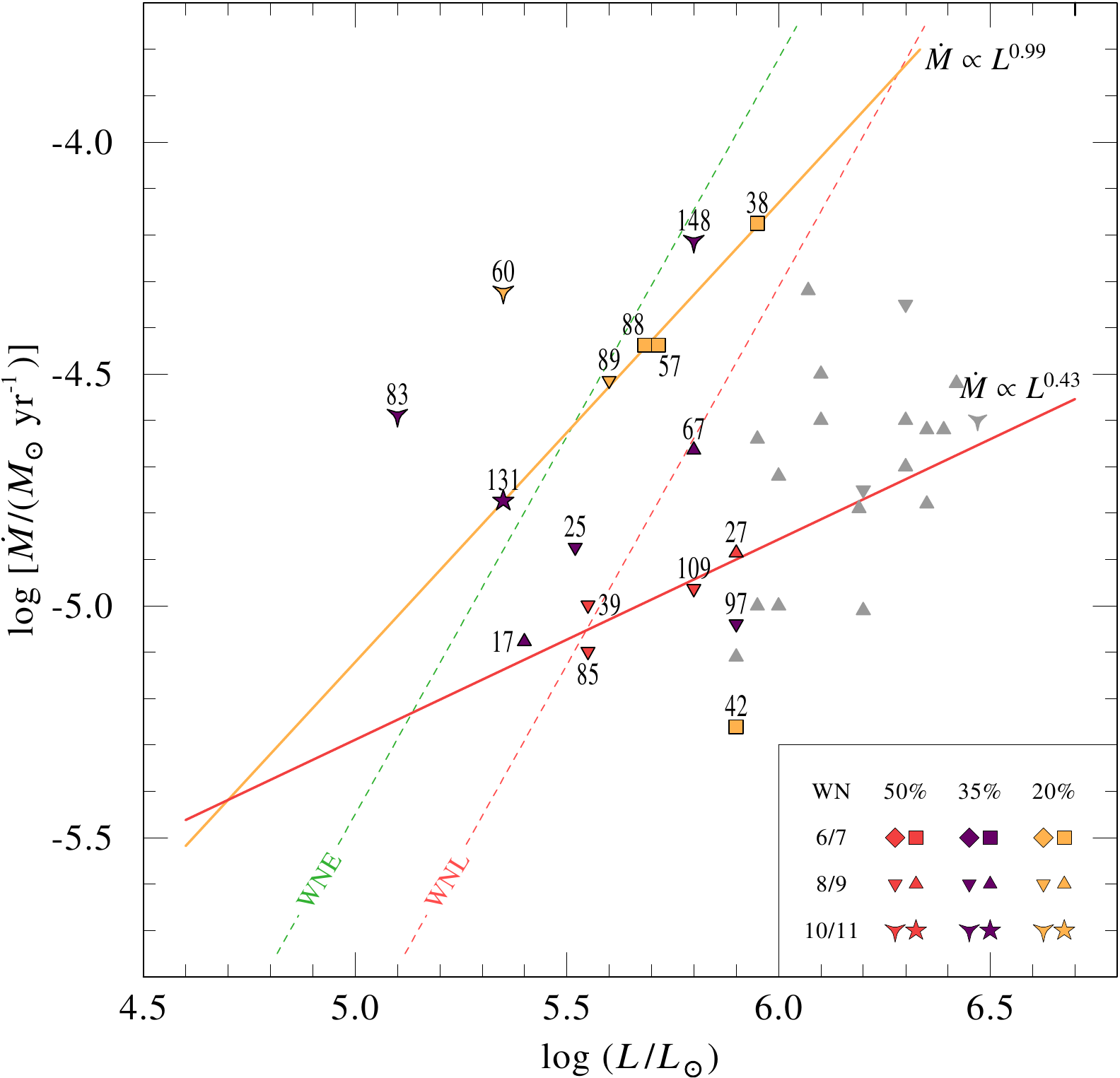}}
  \caption{Empirical mass-loss rate versus luminosity for the M31 late-type 
  				 WN star sample. The thick lines show the least square fits for the 
  				 stars with $X_{\mathrm{H}} \approx 0.2$ and $X_{\mathrm{H}} \approx 0.5$,
  				 respectively. For comparison, the relations from \citet{NL2000} are 
  				 also shown as dashed lines. The gray symbols are the late-type WN 
  				 stars near the Galactic center from \citet{Martins+2008,Liermann+2010}, 
  				 and \citet[][]{Oskinova+2013}.}
  \label{fig:mdot-l}
\end{figure}

  The empirical mass-loss rates are
  plotted in Fig.\,\ref{fig:mdot-l} against the luminosity. There is no significant correlation
  for the whole sample, but one notices a certain trend for the stars with only 20\% hydrogen
  with the exception of the previously discussed \#42 and \#60. There is no subtype sequence 
  visible in the $\dot{M}$-$L$-diagram; instead, the very late WN10 and WN11 are closer to the WN7 stars 
  than most of the WN8 and WN9 stars. This underlines that using the three WN9 to WN11 subtypes, 
  introduced by \citet{CHS1995} instead of the former Ofpe/WN9 classification, is indeed helpful as
  the different subtypes really have different stellar parameters.

%--------- Figure   ----------------------------------------------------
%--------- end Figure ----------------------------------------------------

%--------- Figure   ----------------------------------------------------
\begin{figure}[ht]
  \resizebox{\hsize}{!}{\includegraphics[angle=0]{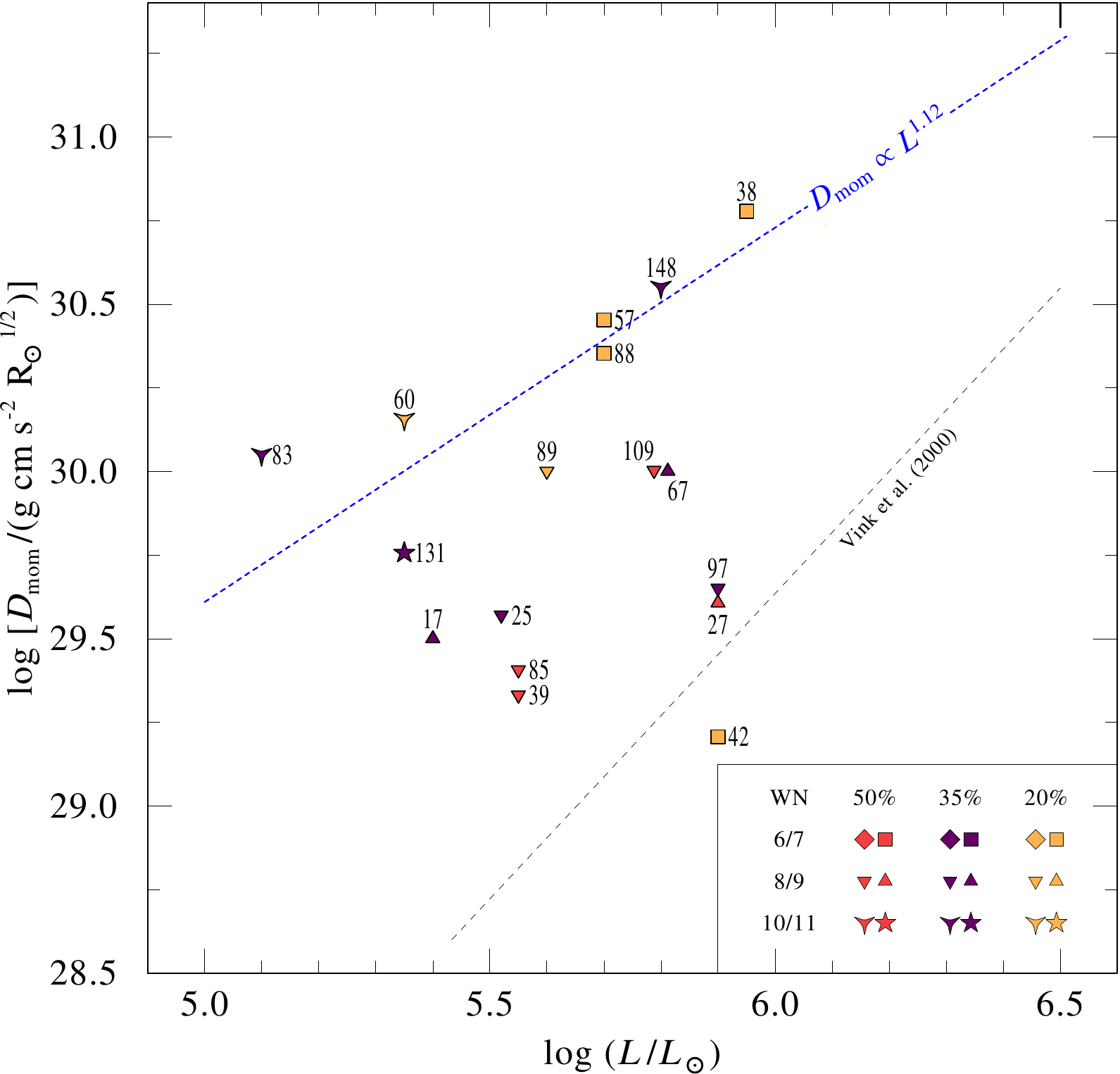}}
  \caption{Modified wind momentum $D_{\mathrm{mom}}$ for the analyzed WN stars, 
  				 plotted against their luminosities. The blue dotted line shows the least 
  				 square fit for the stars with $X_{\mathrm{H}} \leq 0.2$. For comparison, 
  				 the relation from \citet{VdKL2000} for O-stars (black dashed line) is also shown.}
  \label{fig:dmom-l}
\end{figure}
%--------- end Figure ----------------------------------------------------
  
  When comparing these results with the relations suggested by \citet{NL2000}, it becomes
  clear that the mass-loss rates of the WN10 and WN11 stars are higher than expected
  from their luminosities. Interestingly, Galactic WN stars of very late subtype, such as the
  ``peony star'' WR102ka, show the opposite behavior.
  The highest luminosities in our sample are found among the WN8 and WN9 stars.  
  
  In Fig.\,\ref{fig:dmom-l} we plot the so-called \emph{modified wind momentum} 
  $D_{\mathrm{mom}} = \dot{M} \varv_{\infty} \sqrt{R_{\ast}}$ over luminosity.
  Now we can find a nice relation for all those stars with low ($X_{\mathrm{H}} \approx 0.2$) 
  hydrogen content. Such stars are likely to represent the ``classical'' Wolf-Rayet stage with 
  a helium-burning core. For the stars with higher hydrogen content, one cannot distinguish between 
  younger stars evolving to the red side in the Hertzsprung-Russell diagram and slightly older 
  stars evolving back to the blue. Even stars with $X_{\mathrm{H}} \approx 0.2$ could
  in principle still be core-hydrogen burning if they were rapid rotators and therefore
  evolve homogeneously. \citet{Ekstroem+2012} have demonstrated that under this condition, stars with 
  $M_{\text{init}} > 60\,M_{\odot}$ may still be found on the main sequence, despite having only
  20\% of their hydrogen left. Nevertheless, we can discard this scenario as the obtained positions 
  of our WN stars in the Hertzsprung-Russell diagram do not agree with the corresponding evolutionary 
  tracks, which include rotation (see Sect.\,\ref{subsec:tracks}).
  
  The empirical mass-loss rates scale with the square root of the adopted clumping density 
  contrast $D$, for which we assume $D = 4$ \citep{HK1998}. As expected, our $\dot{M}$ results with
  clumped WR models are all higher than predicted by the relation for smooth wind O stars from \citet{VdKL2000}.

%Tabelle mit Sternparametern -----------------------------------------------------------------------------------
\begin{sidewaystable*} 
  \renewcommand{\tabcolsep}{1.5mm}
  \caption{Parameters of the late WN stars from M31}
  \label{tab:fitresults}
  \centering
  \begin{tabular}{r c c c c c c c c c c c c c c c c c c l}
  \hline\hline
    \multicolumn{2}{l}{ M31WR\tablefootmark{a} \hfill	LGGS \rule[0mm]{0mm}{3.5mm}} &	Subtype &  
    \multicolumn{1}{c}{$T_{*}$} & \multicolumn{1}{c}{$\log R_{\mathrm{t}}$} & \multicolumn{1}{c}{$T_{\tau = \frac{2}{3}}$} &
    \multicolumn{1}{c}{$\varv_{\infty}$} & \multicolumn{1}{c}{$E_{B-V}$} & \multicolumn{1}{c}{$M_{V}$} & 
    \multicolumn{1}{c}{$R_{*}$} & \multicolumn{1}{c}{$\log\dot{M}$} & \multicolumn{1}{c}{$\log L$} & 
    \multicolumn{1}{c}{$M_{\varv}$\tablefootmark{b}} & \multicolumn{1}{c}{$\eta$\tablefootmark{c}} & 
    \multicolumn{1}{c}{$\varv_{\mathrm{rad}}^{\mathrm{pec}}$\tablefootmark{d}} & \multicolumn{1}{c}{$M_\text{hom}$\tablefootmark{e}} & \multicolumn{1}{c}{$M_\text{He-b}$\tablefootmark{f}} & \multicolumn{1}{c}{$X_{\mathrm{H}}$} & Comments \\
     & & &
    \multicolumn{1}{c}{[kK]} & \multicolumn{1}{c}{[$R_{\odot}$]} & \multicolumn{1}{c}{[kK]} & 
    \multicolumn{1}{c}{[km/s]} & \multicolumn{1}{c}{[mag]} & \multicolumn{1}{c}{[mag]} &
    \multicolumn{1}{c}{[$R_{\odot}$]} & \multicolumn{1}{c}{[$M_{\odot}/\mathrm{yr}$]} & \multicolumn{1}{c}{[$L_{\odot}$]} &
    \multicolumn{1}{c}{[mag]} & & \multicolumn{1}{c}{[km/s]} & \multicolumn{1}{c}{[$M_{\odot}$]} & \multicolumn{1}{c}{[$M_{\odot}$]} & \multicolumn{1}{c}{[\%]} & \\ 
    \hline \rule[0mm]{0mm}{3.5mm}
17 & J004024.33+405016.2 & WN9 	&  34 	& 1.20 & 32 	& 500 	& 0.12 & -5.62 & 14.1 	& -5.08 & 5.40 & -5.80 & 0.82 & -200 & 27 & 14 & 35 &  \\ 
25 & J004036.76+410104.3 & WN8 	&  50 	& 0.80 & 37 	& 500 	& 0.15 & -5.12 & 7.7 	& -4.87 & 5.52 & -5.40 & 0.99 & -100 & 32 & 16 & 35 &  \\ 
27 & J004056.49+410308.7 & WN9 	&  33 	& 1.20 & 30 	& 300 	& 0.20 & -6.93 & 26.5 	& -4.89 & 5.90 & -7.15 & 0.24 & -200 & 60 & 27 & 50 &  \\ 
38 & J004126.11+411220.0 & WN7 	&  40 	& 0.95 & 33 	& 1000 	& 0.45 & -6.87 & 19.9 	& -4.18 & 5.95 & -6.92 & 3.68 & -150 & 45 & 30 & 20 &  \\ 
39 & J004126.39+411203.5 & WN8 	&  40 	& 0.95 & 31 	& 300 	& 0.55 & -5.98 & 12.6 	& -5.00 & 5.55 & -5.99 & 0.42 & -100 & 39 & 17 & 50 &  \\[2mm] 
42 & J004130.37+410500.9 & WN7-8\tablefootmark{g} 	&  35 	& 1.40 & 35 	& 300 	& 0.22 & -6.58 & 23.7 	& -5.26 & 5.90 & -6.88 & 0.10 & -175 & 42 & 27 & 20 &   +nebula  \\ 
57 & J004238.90+410002.0 & WN7 	&  40 	& 1.00 & 34 	& 1000 	& 0.85 & -6.28 & 14.9 	& -4.44 & 5.70 & -6.24 & 3.58 & -100 & 32 & 21 & 20 &  \\ 
60 & J004242.33+413922.7 & WN10 	&  25 	& 0.80 & 17 	& 300 	& 0.30 & -6.77 & 25.1 	& -4.32 & 5.35 & -6.86 & 3.13 & 0 & 21 & 13 & 20 &    \\ 
67 & J004302.05+413746.7 & WN9 	&  35 	& 1.10 & 31 	& 500 	& 0.75 & -6.58 & 21.1 	& -4.66 & 5.80 & -6.78 & 0.84 & 0 & 44 & 24 & 35 &  \\ 
83 & J004337.10+414237.1 & WN10 	&  25 	& 1.00 & 19 	& 500 	& 0.30 & -5.90 & 18.8 	& -4.59 & 5.10 & -6.09 & 5.03 & 0 & 20 & 9 & 35 &  \\[2mm] 
85 & J004344.48+411142.0 & WN8 	&  35 	& 1.20 & 33 	& 400 	& 0.40 & -6.01 & 15.8 	& -5.10 & 5.55 & -6.10 & 0.44 & 0 & 39 & 17 & 50 &   +nebula \\ 
88 & J004353.34+414638.9 & WN7 	&  50 	& 0.80 & 38 	& 1000 	& 0.10 & -5.74 & 9.4 	& -4.44 & 5.70 & -5.81 & 3.58 & 0 & 32 & 21 & 20 &  \\ 
89 & J004357.31+414846.2 & WN8 	&  45 	& 0.70 & 30 	& 500 	& 0.70 & -6.03 & 10.6 	& -4.51 & 5.60 & -6.09 & 1.89 & 0 & 28 & 18 & 20 &  \\ 
97 & J004413.06+411920.5 & WN8 	&  35 	& 1.40 & 35 	& 500 	& 0.35 & -6.58 & 23.7 	& -5.04 & 5.90 & -6.89 & 0.28 & 0 & 50 & 27 & 35 &  \\ 
109 & J004430.04+415237.1 & WN8 	&  35 	& 1.50 & 35 	& 1000 	& 0.45 & -6.53 & 21.1 	& -4.96 & 5.80 & -6.64 & 0.85 & 200 & 53 & 24 & 50 &   diluted? \\[2mm] 
131 & J004511.21+420521.7 & WN11 	&  22 	& 1.20 & 20 	& 300 	& 0.30 & -6.69 & 31.5 	& -4.77 & 5.35 & -6.87 & 1.11 & 100 & 26 & 13 & 35 &  \\ 
148 & J004542.26+414510.1 & WN10 	&  28 	& 1.00 & 22 	& 500 	& 0.75 & -7.38 & 33.4 	& -4.21 & 5.80 & -7.52 & 2.38 & 100 & 44 & 24 & 35 &  \\ 
  \hline\end{tabular}
  \tablefoot{\\
  	 \tablefoottext{a}{The M31WR numbers correspond to the sequence of entries in Table 5 from \citet{NMG2012} which is sorted
  	                   by right ascension.\\}
  	 \tablefoottext{b}{Absolute monochromatic magnitude $M_{\varv}$ as defined by \citet{S1968}, reconstructed with our model SED.\\}
     \tablefoottext{c}{The wind efficiency parameter $\eta$ is defined as the ratio of wind and photon momentum: $\eta := \frac{\dot{M} \varv_{\infty} c}{L}$\\}
  	 \tablefoottext{d}{Peculiar radial velocity, $\varv_{\mathrm{rad}}^{\mathrm{pec}}$, describing the radial velocity of the star in addition to the mean radial velocity of M31 ($-301\,$km\,s$^{-1}$)\\}
     \tablefoottext{e}{Current masses calculated from luminosity via the relation from \citet{Graefener+2011}, which takes the different
                       hydrogen fractions into account. These masses are based on the assumptions of a chemically homogeneous
                       star.\\}
     \tablefoottext{f}{Current mass calculated from luminosity via \citet{Graefener+2011} for a helium-burning star with a negligible mass of the hydrogen-containing outer layers.\\}
     \tablefoottext{g}{Own classification}
  }  
\end{sidewaystable*}
%-------------------------------------------------------------------------------------------------------------

\subsection{Hertzsprung-Russell diagram}
  \label{subsec:hrd}

%--------- Figure   ----------------------------------------------------
\begin{figure}[ht]
  \resizebox{\hsize}{!}{\includegraphics[angle=0]{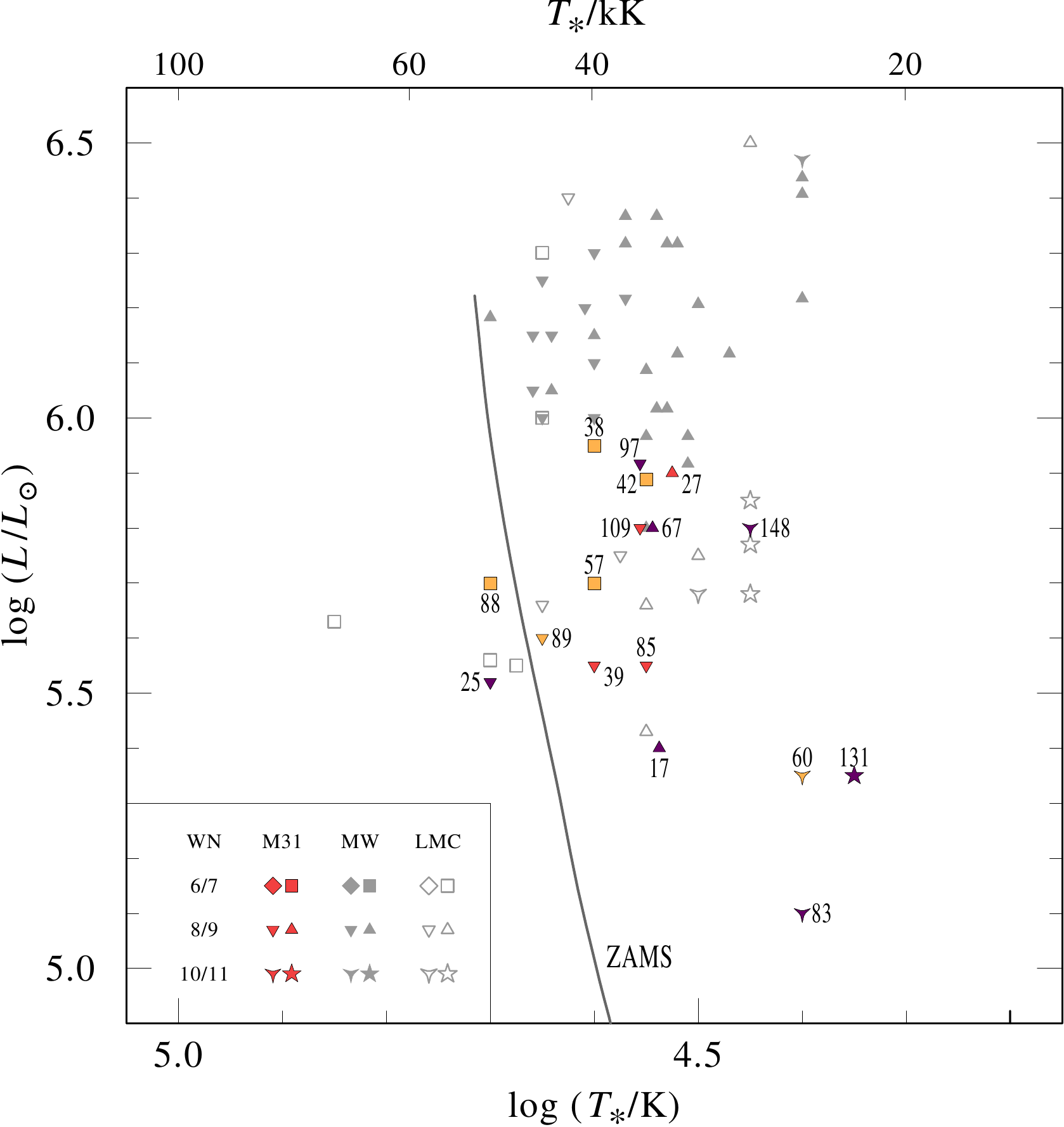}}
  \caption{Hertzsprung-Russell diagram of the M31 late WN sample. 
  				 The colored symbols refer to the M31 stars with the colors indicating different
  				 hydrogen content as labeled in the previous figures. The filled gray symbols show 
  				 the positions of the late-type WNh stars from Milky Way using the results from \citet{HGL2006},
  				 \citet{Martins+2008} and \citet{Liermann+2010}. The open gray symbols refer to
  				 the LMC late-type WNh stars from \citet{Hainich+2014}.}
  \label{fig:hrd-wncmp}
\end{figure}
%--------- end Figure ----------------------------------------------------

  The Hertzsprung-Russell diagram of the sample is shown in Fig.\,\ref{fig:hrd-wncmp} with
  the hydrogen fraction color-coded and the different symbol shapes indicating the
  WN subtypes. For comparison, the late-type WN stars from the Galaxy \citep{HGL2006,Martins+2008,Liermann+2010}
  and the LMC \citep{Hainich+2014} are plotted. Almost all stars are located on the cool side of the zero-age main sequence.
  
  There are no M31 stars above $\log L/L_{\odot} = 6.0$, in contrast to several Galactic WN stars 
  and at least a few LMC stars. While a part of the
  luminosities in the Galactic sample might suffer from unsure distances, the sample also includes 
  several stars from the Galactic center region, which is thought to be quite well known in terms
  of distance. Instead of resembling the Milky Way positions in the HRD, the M31 results seem
  to be closer to what we see for the LMC WN population, despite the major differences between 
  these two galaxies. 
  
  No star in our sample is from the innermost part of M31, and
  it is exactly this part of the Milky Way where we find the WN stars with the highest 
  luminosities, such as the so-called ``Peony nebula star'' WR102ka \citep[][]{BOH2008,Oskinova+2013}. 
  Therefore the picture might change once we get access to late-type WN stars
  in this innermost part. 

\subsection{Subtype-magnitude relation}
  \label{subsec:mvrel}

  Using the spectral energy distribution from the best-fitting model atmospheres for each
  star, we can obtain the monochromatic magnitude $M_{\varv}$ defined by \citet{S1968}. While
  the individual value for each object are listed in Table\,\ref{tab:fitresults}, the arithmetic
  mean and the standard deviation for each subtype are given in Table\,\ref{tab:mvmean}. All stars 
  that entered this calculation contain hydrogen, although the exact amount is different. Due to the
  problem that the distances are not known for several Galactic WR stars, the only way was to 
  adopt a subtype-magnitude calibration in order to obtain the luminosities.
 
%--------- Figure   ----------------------------------------------------
\begin{figure}[ht]
  \resizebox{\hsize}{!}{\includegraphics[angle=0]{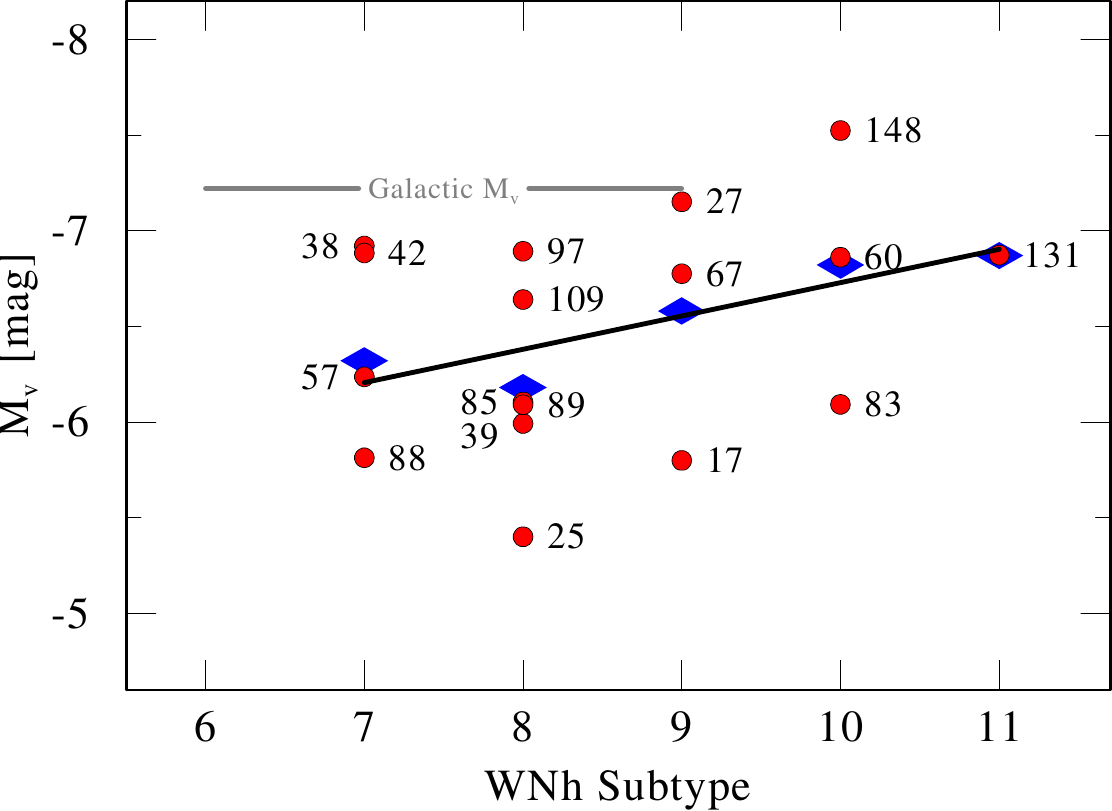}}
  \caption{Mean absolute magnitudes of the WN stars in M31 plotted per 
           subtype. The red dots indicate the positions of the individual
           stars while the blue diamonds mark the mean value per subtype.
           The black line shows the linear fit. For
           comparison, the Galactic relation used in \citet{HGL2006} is 
           shown in gray.}
  \label{fig:mvstats}
\end{figure}
%--------- end Figure ----------------------------------------------------

\begin{table}
  \caption{Mean absolute magnitudes per WN subtype in M31}
  \label{tab:mvmean}
  \centering
  \begin{tabular}{l c l l}
  \hline\hline
    Subtype &  \multicolumn{1}{c}{$M_{\varv}$ [mag]} & \multicolumn{2}{r}{$\Delta M_{\varv}$ [mag]\tablefootmark{a}} \rule[0mm]{0mm}{3.5mm} \\
  \hline
    WN7 & -6.32 & \rule[0mm]{0mm}{3.5mm} & 0.46 \\
    WN8 & -6.18 &  & 0.48 \\
    WN9 & -6.58 &  & 0.57 \\
    WN10 & -6.82 &  & 0.58 \\
    WN11 & -6.87 &  & 0.00\tablefootmark{b} \\
  \hline
  \end{tabular}
  \tablefoot{
     \tablefoottext{a}{standard deviation}
     \tablefoottext{b}{only one star available}
  }
\end{table}

  The results from M31 now allow us to check the validity of such a calibration for late-type
  WN stars. \citet{HGL2006} assumed a brightness magnitude of $M_{\varv} = -7.22\,$mag for all late-type WN
  stars with hydrogen. With the known distance of M31, we obtain a mean magnitude of $M_{\varv} = -6.47\,$mag for
  our sample. However, if we plot the individual $M_{\varv}$ versus the subtype (Fig.\,\ref{fig:mvstats}),
  a substantial scatter is encountered. This probably reflects the fact that WN stars do not
  form a one-parameter sequence, since they differ not only in mass, but also in their inner
  structure. Thus the magnitude calibration cannot work precisely and an uncertainty of about
  $\pm0.5\,$mag remains. Moreover, Fig.\,\ref{fig:mvstats} indicates a systematic trend of $M_{\varv}$
  with subtype rather than a constant value, as reflected by the shown linear regression fit.
  
  After the LMC analyses \citep{Hainich+2014}, the present results are a further 
  indicator that at least a part of the Galactic WN stars might be less distant and therefore 
  less luminous than assumed hitherto. 
   
\subsection{Error estimates}
  \label{subsec:errors}
  
  The careful fitting process and the calculation of additional models in those cases where
  a grid model did not reproduce the observed spectrum well enough meant that the uncertainties in the two
  main parameters $T_{\ast}$ and $R_{\mathrm{t}}$ are below one grid cell, i.e., $\leq 0.05\,$dex in 
  $T_{\ast}$ and $\leq 0.1\,$dex in $R_{\mathrm{t}}$. For the terminal velocity $\varv_{\infty}$
  the uncertainty is approximately 25\%. The luminosities not only depend on the models, but also on the accuracy
  of the available photometry, which is only broadband. The distance to M31 is a minor source of 
  uncertainty. In total, we assume a rather conservative error of $0.1\,$dex in the stellar luminosity $L$. 
  According to Eq.\,(\ref{eq:rt}), the uncertainty propagates to $0.15\,$dex in the mass-loss rate $\dot{M}$.
  With this value given, the error for the modified wind momentum $D_{\mathrm{mom}}$ is approximately $0.3\,$dex.
  These error margins do not include systematic errors from the various 
  assumptions and atomic data used by the stellar atmosphere models that are impossible to quantify.
  
\section{Discussion of the evolutionary status}
  \label{sec:evol}  

\subsection{Late WN stars as LBV candidates}
  \label{subsec:wnlbv}

  The spectra of late WN stars are very similar to those of luminous blue variables (LBVs).
  In fact, some known LBVs have shown a WN spectral type during their quiet stage, such as 
  AG\,Car in the late 1980s \citep{Stahl+2001}. The term LBV was introduced 
  by \citet{C1984} for hot, luminous, and variable stars
  that do not fit into the WR class, such as $\eta\,$Car, P\,Cyg, and S\,Dor. 
  The first two of the three stars were already known as variable objects in the 16th century,
  but it was not until the 1970s when their similarities and their possible connection to a certain 
  stage in the massive star evolution was discussed \citep[see e.g.][]{H1975,S1975,H1978,HD1979,WAC1980}.
  
  Today the term LBV is used incoherently in the literature. Although the name might 
  suggest that every luminous and variable blue star
  could be called an LBV, the original idea was to define this class by a 
  particular type of variability, including phases when the LBV is currently
  not ``blue'' \citep{HD1994}. This variability is not theoretically 
  understood, but observed for several objects and named after the prototypical star S\,Doradus.
  In quiescence these stars occupy a certain range in the HR diagram called the S\,Dor
  instability strip \citep{W1989}. During ``normal outbursts'' they brighten by one to two magnitudes
  in the optical while their bolometric magnitude does not change much.
  
	Combining earlier works such as \citet{HD1994} with newer results, \citet{vG2001} sets
	up a list of criteria that qualify a star as an LBV candidate. Instead of the vague term 
	LBV he speaks of S Dor (SD) variability and SD-membership which needs to fulfill three criteria, 
	namely photometric variability, visible ejecta, and spectroscopic characteristics that 
	indicate a high luminosity and mass loss, together with the presence of CNO-processed material. Consequently, 
	WN9 to WN11 stars are not automatically considered as LBV candidates unless they show stellar ejecta. 
	If a star has a spectrum that resembles those of a known LBV,
	and it is known that both brightness and spectral appearance changed significantly on
	timescales of a few years, it is called a \emph{bona fide} LBV.
	
%--------- Figure   ----------------------------------------------------
\begin{figure}[ht]
  \resizebox{\hsize}{!}{\includegraphics[angle=0]{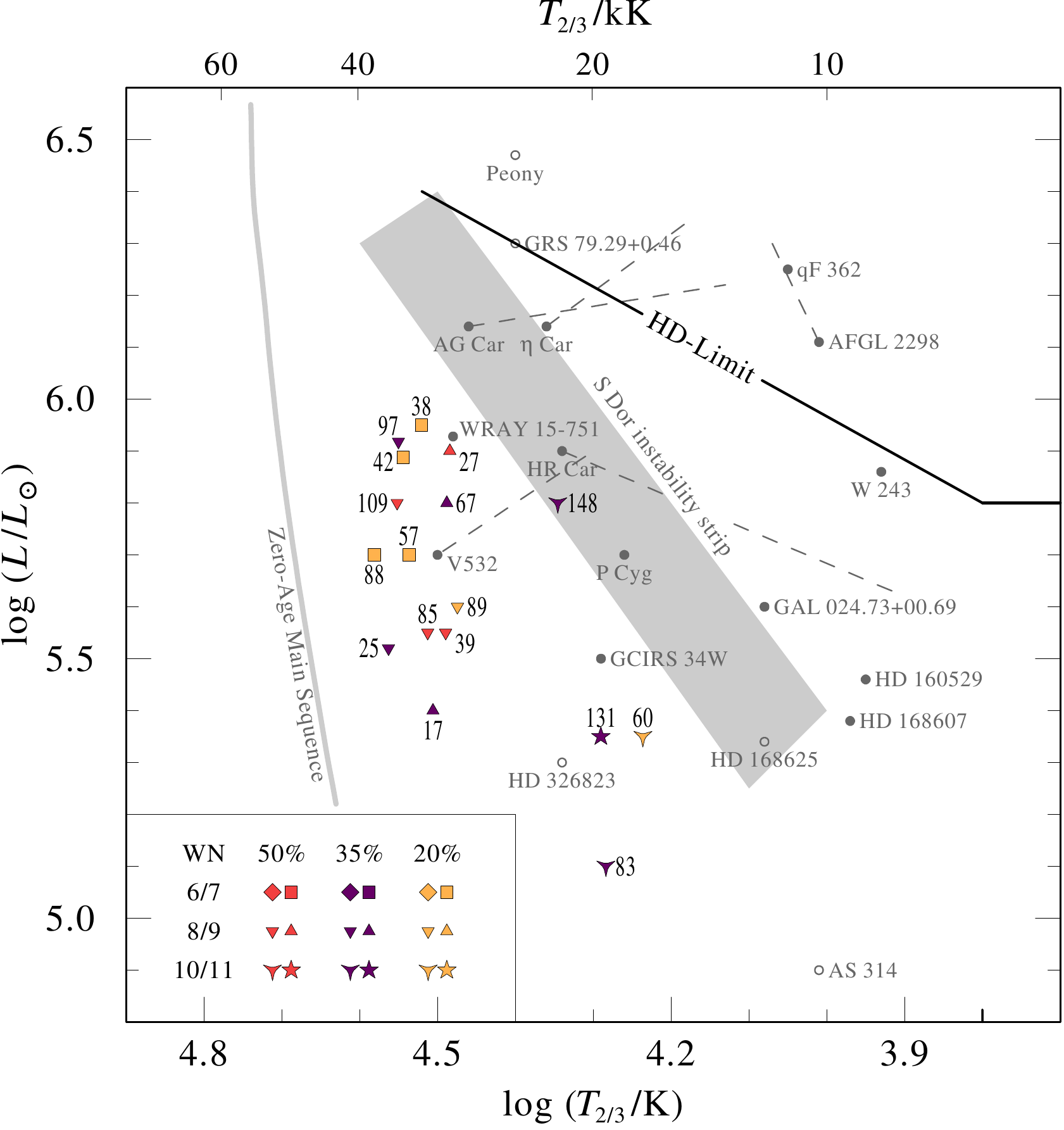}}
  \caption{Hertzsprung-Russell diagram based of the M31 
           late-WN sample, this time with $T_{\tau = \frac{2}{3}}$ instead of 
           $T_{\ast}$. For comparison, known LBVs (filled circles) and LBV candidates
           (open circles) are shown.
           The gray-shaded area marks the S Dor instability strip \citep{W1989}. The
           empirical Humphreys-Davidson limit is also indicated. The parameters of the comparison 
           stars are taken from \citet{NRH2012} with the exception of
           Romano`s Star \citep[V532,][]{CCG2012,MA2012} and the Peony star \citep{BOH2008}. The
           locations of LBVs in different stages are connected by dashed lines. Labels
           without letters refer to the M31WR numbers as listed in Table\,\ref{tab:fitresults}.}
  \label{fig:hrd-lbv}
\end{figure}
%--------- end Figure ----------------------------------------------------

  Figure\,\ref{fig:hrd-lbv} displays the positions of the analyzed stars in the 
  Hertzsprung-Russell diagram (HRD) in comparison to other known LBVs and 
  LBV candidates. In contrast to the version shown in Fig.\,\ref{fig:hrd-wncmp},
  we now use as the abscissa the effective temperature $T_{\tau = 2/3}$, which refers to the radius
  where the Rosseland optical depth reaches two thirds. Remember that $T_{\ast}$, which is used in
  Fig.\,\ref{fig:hrd-wncmp}, refers to the radius $R_{\ast}$ at $\tau_{\text{Ross}} = 20$, which might be
  close to the radius of the hydrostatic core. None of the M31 stars is located
  beyond the empirical Humphreys-Davidson limit \citep{HD1979}, and only one star (\#148)
  is in the region of the so-called S Dor instability strip found by \citet{W1989}.
  However, several of our sample stars are not far from this region, while some known LBVs like WRAY\,15-751
  are close to it, but outside. The bona fide LBV V532, known as Romano`s star, even
  appears as WN8 during its hot phase \citep{CCG2012,MA2012}. We therefore cannot rule out that at 
  least a subsample of our analyzed stars, especially those located at a very similar position as known LBVs, such as \#38 or 
  \#27, could represent LBVs during their quiet stage. In all galaxies where the WR population
  has been studies comprehensively (MW, LMC, M31), the very late WN10 and WN11 subtypes are extremely rare. 
  Although there is a certain 
  selection bias in the finding methods as mentioned in Sect.\,\ref{subsec:name}, this might
  indicate that a star appearing as WN10-11 will not remain in this state for a long time.
   
  One star in our sample,	\#60, was mentioned as a P\,Cygni-like LBV candidate 
  by \citet{Massey+2007} but later listed as very late WN (WNVL) in \citet{NMG2012}. The star J004334.50+410951.7,
  classified as Ofpe/WN9 in \citet{Massey+2007}, does not appear in the M31 WR census of \citet{NMG2012}.

\subsection{Comparison with evolutionary tracks}
  \label{subsec:tracks}

  The evolution of massive stars in the upper part of the HRD is still poorly understood.
  It is not really clear which initial mass ranges lead to the different kinds of WR types
  and how this depends on the galactic environment and metallicity. Recent studies by
  \citet{MP2013} have shown that different evolutionary codes predict
  significantly different results, especially for hydrogen-rich WN stars above $\log L/L_{\odot} > 5.3$.
  
	To discuss the evolutionary situation for the analyzed WN sample, 
	the empirically obtained positions in the HRD are compared to evolutionary tracks. In the
	past years it has become clear that at least a fraction of the hydrogen-rich WN stars are 
	still core-hydrogen burning \citep[see, e.g.,][]{Graefener+2011}. In fact, the most massive stars currently known
	are hydrogen-rich WN stars \citep[][]{Crowther+2010,Hainich+2014}. 
	
\begin{figure}[ht]
  \resizebox{\hsize}{!}{\includegraphics[angle=0]{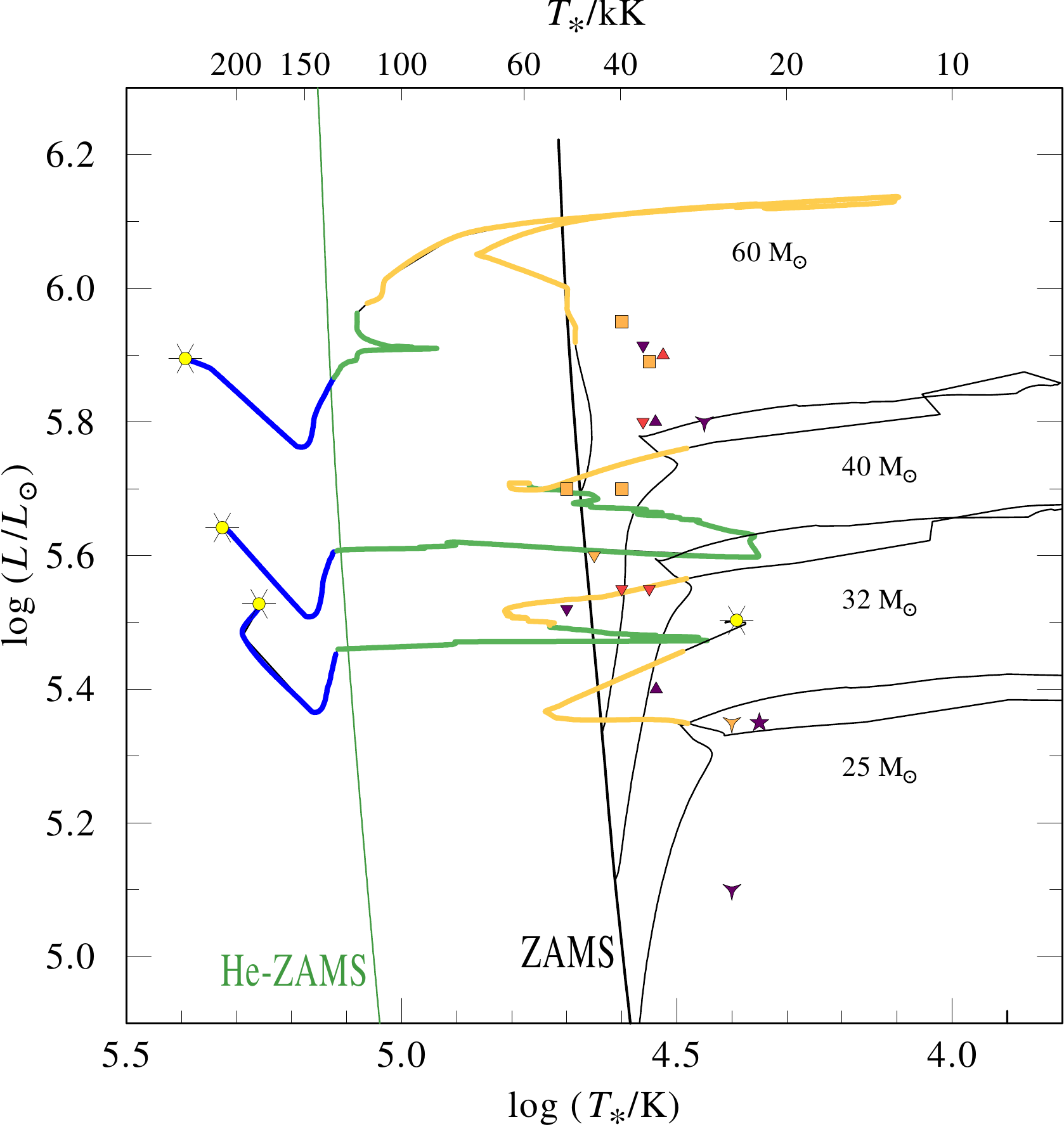}}
  \caption{HR diagram of the M31 late WN sample compared with the latest set
           of Geneva evolutionary tracks for $Z=0.014$ (with rotation) from \citet{Georgy+2012}.
           The thick orange part of the tracks corresponds to the 
           hydrogen-rich WN phase, while the green and the blue parts indicate 
           the hydrogen-free WN and the WC (incl. WO) phase, respectively.}
  \label{fig:hrd-geneva-rot}
\end{figure}

  In Fig.\,\ref{fig:hrd-geneva-rot}, we compare our sample to	the current Geneva 
  evolutionary tracks for $Z=0.014$ from \citet{Georgy+2012}, which include rotation. 
	Based on the surface abundances predicted by the evolutionary models, we highlighted
	the different WR phases of the tracks using different colors: The hydrogen-rich 
	($X_\text{H} > 0.05$) WN stage, the hydrogen-free ($X_\text{H} < 0.05$) WN stage, and 
	the WC stage, which we infer from a carbon surface fraction $X_\text{C} \geq 0.2$.
	
\begin{figure}[ht]
  \resizebox{\hsize}{!}{\includegraphics[angle=0]{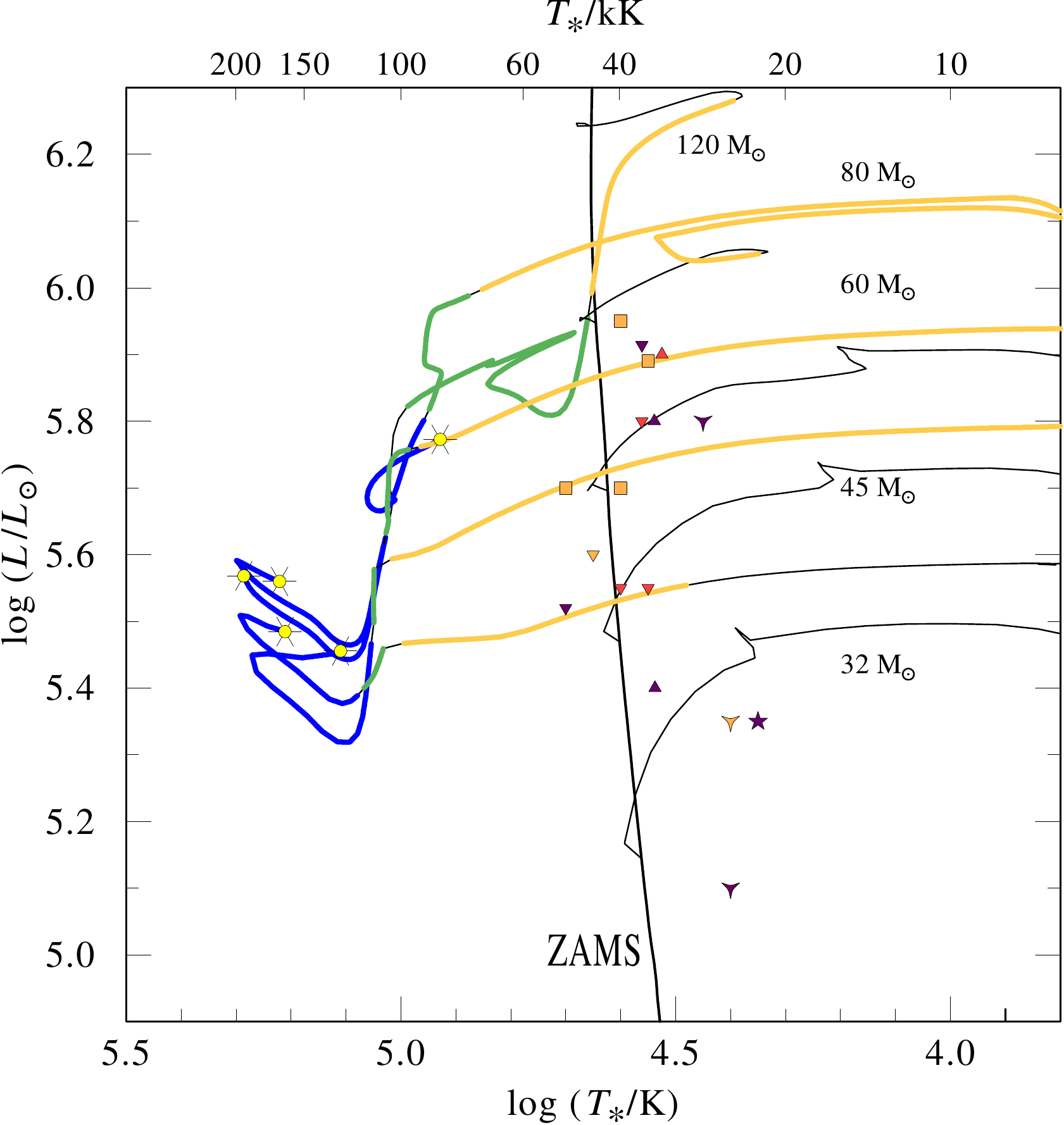}}
  \caption{HR diagram of the M31 late WN sample compared with the single
  				 star evolutionary tracks for $Z=0.02$ from \citet{EIT2008}. The
  				 track colors have the same meaning as in Fig.\,\ref{fig:hrd-geneva-rot}.}
  \label{fig:hrd-eldridge}
\end{figure}

\begin{figure}[ht]
  \resizebox{\hsize}{!}{\includegraphics[angle=0]{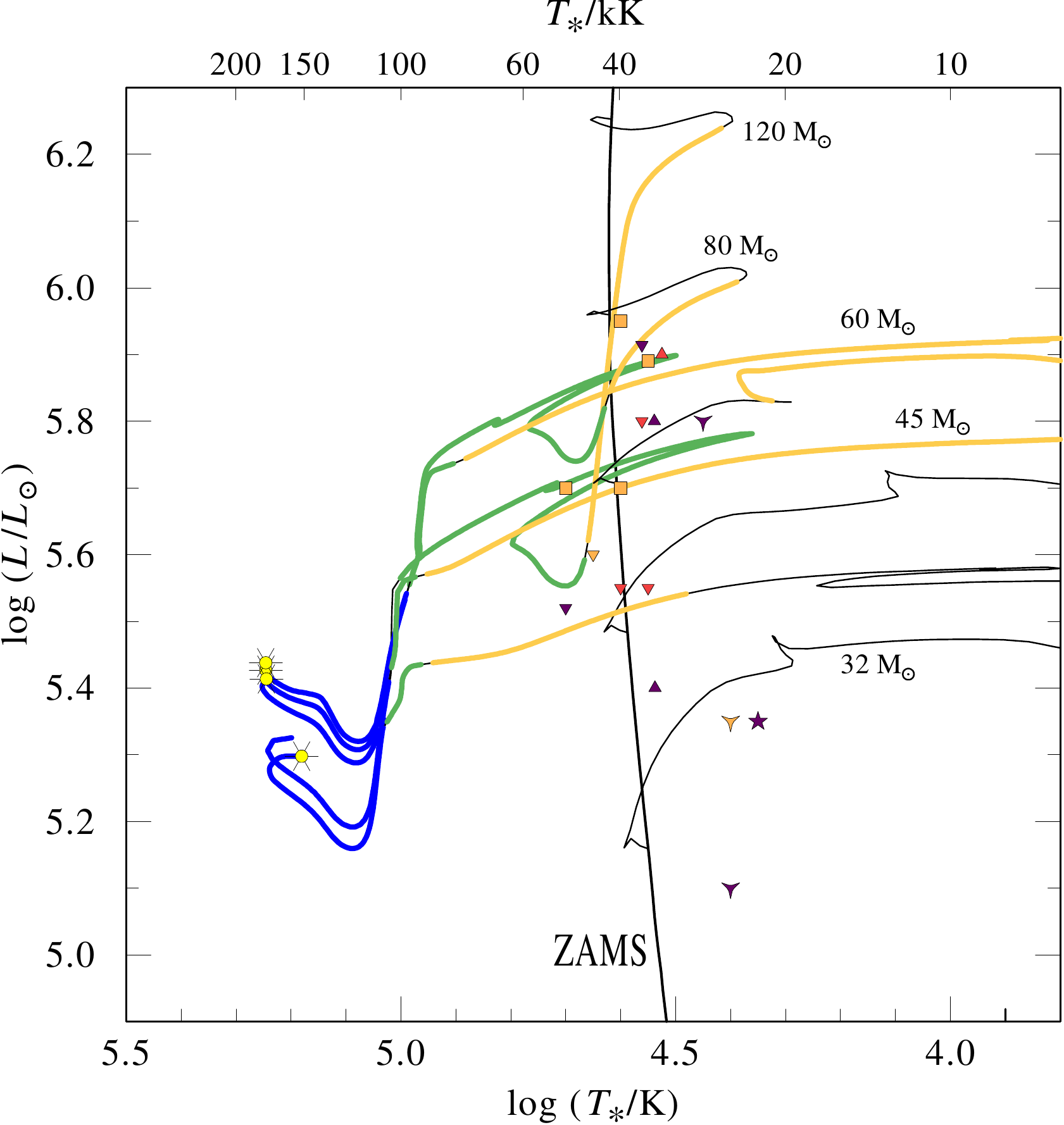}}
  \caption{Same as Fig.\,\ref{fig:hrd-eldridge}, but for $Z=0.03$.}
  \label{fig:hrd-eldridge-z030}
\end{figure}

	For the majority of the stars in our sample there is good agreement with the 
	hydrogen-rich WN phase of the Geneva tracks. 
	However, the tracks cannot explain the low luminosities of the WN11 star \#131 and the two
	WN10 stars \#60 and \#83. Even if we take into account that some regions of M31
	have a higher metallicity than our Galaxy and therefore compare them with tracks for $Z = 0.02$
	and $Z = 0.03$ from \citet{EIT2008} as shown in Figs.\,\ref{fig:hrd-eldridge} and \ref{fig:hrd-eldridge-z030}, 
	we do not find any track that predicts WR stars at these positions.
	  
%--------- Figure   ----------------------------------------------------
%--------- end Figure ----------------------------------------------------

%--------- Figure   ----------------------------------------------------
%--------- end Figure ----------------------------------------------------

%--------- Figure   ----------------------------------------------------
%--------- end Figure ----------------------------------------------------
  
  Both sets of tracks, the Geneva tracks with rotation and the Eldridge tracks without 
  rotation (Figs.\,\ref{fig:hrd-geneva-rot} and \ref{fig:hrd-eldridge}, respectively), fit reasonably well with the 
  HRD positions of the analyzed sample. Apart from the objects mentioned in the
  previous paragraph, other stars with lower luminosities, such as \#39 or \#75
  could be explained either by a 40\,$M_{\odot}$-track without initial rotation or
  a 32\,$M_{\odot}$-track with an initial rotation of $400\,$km/s. The stars 
  \#60 or \#131 only match the tracks accounting for rotation. 
  
  The WN stars analyzed in this paper represent an early stage of the WR evolution
  where the evolutionary tracks are not very sensitive to details such as rotation.
  In later stages they differ more. As a result, our forthcoming analyses of the early-type 
  WN stars in M31 will give us more information about which set of tracks is more adequate.
  
  So far, we can conclude that the known late-type WN stars in M31 stem
  from an initial mass range between $20$ and $\approx 60\,M_{\odot}$, if formed via single-star
  evolution. When the objects with the lowest luminosity have undergone
  binary evolution, the lower limit would rise to $25$ or even $30\,M_{\odot}$,
  depending on how many stars are discarded and which set of tracks is appropriate.
  By using binary models from \citet{Eldridge+2013}, we can indeed explain
  the HRD position of \#83, the star with the lowest luminosity in our sample, as
  the product of binary evolution with an initial mass of $\approx 20\,M_{\odot}$.
  
  The highest initial mass in our sample depends even more on the stellar evolution models. The Geneva tracks 
  (Fig.\,\ref{fig:hrd-geneva-rot}) indicate that the most luminous star in our sample (\#38) had an initial
  mass below $60\,M_{\odot}$ while the tracks from \citet{EIT2008} for $Z = 0.02$ (Fig.\,\ref{fig:hrd-eldridge}) 
  would indicate an initial mass above $60\,M_{\odot}$. The tracks for $Z = 0.03$ (Fig.\,\ref{fig:hrd-eldridge-z030})
  would increase this value even to $80\,M_{\odot}$, but this higher metallicity is not
  supported by our spectral fits, as is evident from the good match of the nitrogen lines (cf. Fig.\,\ref{fig:m31wr38}).
  
  For none of the WR stars in M31, we obtained a luminosity of $\log L/L_{\odot} > 6.0$,
  which would indicate a very massive star. Interestingly, this situation does not seem
  to be restricted to WR stars. A current work from \citet{Humphreys+2013} dealing
  with luminous and variable stars does not list any star in M31 above this limit either.  

  However, the most massive stars so far have been found in very massive clusters or
  special environments, such as the central region of our Galaxy or R136 in the LMC.
  It might just be a selection effect that such stars have not been identified M31 so far.
  On the other hand, if there really are no very massive stars in M31, this 
  would match with the low SFR calculated by \citet{NMG2012}.
  
\section{Conclusions}
  \label{sec:conclusions}
  
  For the first time, we have analyzed a complete sample of late-type WN stars in a 
  Milky-Way-like galaxy. We analyzed 17 late-type WN stars in the Andromeda Galaxy (M31), 
  including all known WN7 to WN11 stars with available spectra of sufficient quality. With 
  the known distance of M31, we can avoid the large uncertainty in the luminosities of 
  Galactic WR stars. We draw the following conclusions:
  
  \begin{itemize}
    \item All stars in our sample have luminosities $L$ between $10^5$ and $10^6\,L_{\odot}$. 
          Notably, we do not find any star in our sample that exceeds $10^6\,L_{\odot}$.
    \item The absolute visual magnitude $M_{\text{V}}$ shows a significant scatter,
          even within the same WN subtype. This sets limits to the applicability of a subtype 
          magnitude calibration. 
    \item Since the current catalog of WN stars in M31 does not
          include the central region of this galaxy, we cannot rule out that
          this special environment hosts WN stars with luminosities $L > 10^6\,L_{\odot}$, 
          especially since we know that such stars exist in the LMC and the Galactic center region. 
    \item The late-type WN stars in M31 are in good agreement with the latest set of the Geneva
          tracks with $Z = 0.014$. Only the Geneva tracks with rotation can reproduce the lowest 
          luminosities in our sample.
          Otherwise, the different sets of tracks \citep{Georgy+2012,EIT2008} can all explain the
          observed WNh stars because these stars are not yet in the final WR stages for which the tracks
          differ more. 
    \item If formed via single-star evolution, the analyzed late-type WN stars in M31 stem from an
          initial mass range between $20$ and $60\,M_{\odot}$. The lower limit might in fact be
          higher if the few stars with the lowest luminosities were formed via binary evolution.
    \item Our spectral fits reveal that the analyzed late-type WN stars have similar chemical 
    			compositions to their Galactic counterparts. This is in line with the fact that the stars 
    			are not located in the inner part of M31, where we would expect higher metallicities.
    			The obtained HRD positions of the analyzed
    			stars do not provide additional constraints since they can be reproduced by evolutionary 
    			models with both Galactic and higher metallicity.
    \item The number of very late WN stars (WN9-11) is low, which could partly be a selection
          effect of the detection method \citep{NMG2012}. While a lot of additional WN9 stars have been 
          discovered in the Milky Way in obscured regions with infrared observations, the number
          of currently known WN10 and WN11 stars in M31 is about the same as in the Milky Way and 
          the LMC.
    \item Only one object, the WN10 star \#148, is located in the S\,Dor instability strip
          in the HR-diagram.
          However, the HRD positions do not rule out that some of the stars of very late-type WN subtypes might
          be LBVs in the quiescent stage. From the luminosity range, it is hard to tell whether
          the WN stars have lost parts of their hydrogen via an LBV outbursts or in the red supergiant
          (RSG) phase.
  \end{itemize}
  
  In a forthcoming paper we will analyze the early WN subtypes to get a more complete 
  picture of the WN population in M31.

\begin{acknowledgements}
  We would like to thank the referee, Cyril Georgy, for his constructive feedback
  that improved the present work. This research has made use of the SIMBAD database and the VizieR 
  catalog access tool, both operated at the CDS, Strasbourg, France. The Digitized Sky Surveys 
  were produced at the Space Telescope Science Institute under U.S. Government grant NAG W-2166. 
  The images of these surveys are based on photographic data obtained using the Oschin Schmidt 
  Telescope on Palomar Mountain and the UK Schmidt Telescope. The plates were processed into 
  the present compressed digital form with the permission of these institutions. We acknowledge 
  the use of NASA's SkyView facility located at NASA Goddard Space Flight Center. The first author 
  of this work (A.S.) is supported by the Deutsche Forschungsgemeinschaft (DFG) under grant HA 1455/22.
\end{acknowledgements}

%- - - - - - - - - - - - - - - - - - - - - - - - - - - - - - - - - - - - - - - - - - - - - - 

% for the bibliography, at the end
\bibliographystyle{aa} % style aa.bst
\bibliography{m31wnlpaper}

%------------------- ONLINE MATERIAL --------------
%\end{document}

\Online
\label{onlinematerial}

\begin{appendix} 

\section{Spectral Fits}
  \label{appsec:specfits}

\begin{table}[h!]
  \renewcommand{\tabcolsep}{1.5mm}
  \caption{Appendix overview: Analyzed late-type WN stars in M31}
  \label{tab:onlineindex}
  \centering

  \begin{tabular}{c l l l l}
    \hline %-------------------------------------------------------------
    \hline %-------------------------------------------------------------
    M31WR &  LGGS  &  WR type  &  Figure  &  Page \rule[0mm]{0mm}{3mm}  \\
    \hline %-------------------------------------------------------------
17  & J004024.33+405016.2 & WN9 &   \ref{fig:m31wr17}  & \pageref{fig:m31wr17} \\
25  & J004036.76+410104.3 & WN8 &   \ref{fig:m31wr25}  & \pageref{fig:m31wr25} \\
27  & J004056.49+410308.7 & WN9 &   \ref{fig:m31wr27}  & \pageref{fig:m31wr27} \\
38  & J004126.11+411220.0 & WN7 &   \ref{fig:m31wr38}  & \pageref{fig:m31wr38} \\
39  & J004126.39+411203.5 & WN8 &   \ref{fig:m31wr39}  & \pageref{fig:m31wr39} \\[2mm]
42  & J004130.37+410500.9 & WN7-8 & \ref{fig:m31wr42}  & \pageref{fig:m31wr42} \\
57  & J004238.90+410002.0 & WN7 &   \ref{fig:m31wr57}  & \pageref{fig:m31wr57} \\
60  & J004242.33+413922.7 & WN10 &  \ref{fig:m31wr60}  & \pageref{fig:m31wr60} \\
67  & J004302.05+413746.7 & WN9 &   \ref{fig:m31wr67}  & \pageref{fig:m31wr67} \\
83  & J004337.10+414237.1 & WN10 &  \ref{fig:m31wr83}  & \pageref{fig:m31wr83} \\[2mm]
85  & J004344.48+411142.0 & WN8 &   \ref{fig:m31wr85}  & \pageref{fig:m31wr85} \\
88  & J004353.34+414638.9 & WN7 &   \ref{fig:m31wr88}  & \pageref{fig:m31wr88} \\
89  & J004357.31+414846.2 & WN8 &   \ref{fig:m31wr89}  & \pageref{fig:m31wr89} \\
97  & J004413.06+411920.5 & WN8 &   \ref{fig:m31wr97}  & \pageref{fig:m31wr97} \\
109  & J004430.04+415237.1 & WN8 &  \ref{fig:m31wr109} & \pageref{fig:m31wr109} \\[2mm]
131  & J004511.21+420521.7 & WN11 & \ref{fig:m31wr131} & \pageref{fig:m31wr131} \\
148  & J004542.26+414510.1 & WN10 & \ref{fig:m31wr148} & \pageref{fig:m31wr148} \\
    \hline %-----------------------------------------------------
  \end{tabular}
\end{table}

\clearpage

\begin{figure*}
  \centering
  \includegraphics[width=14cm]{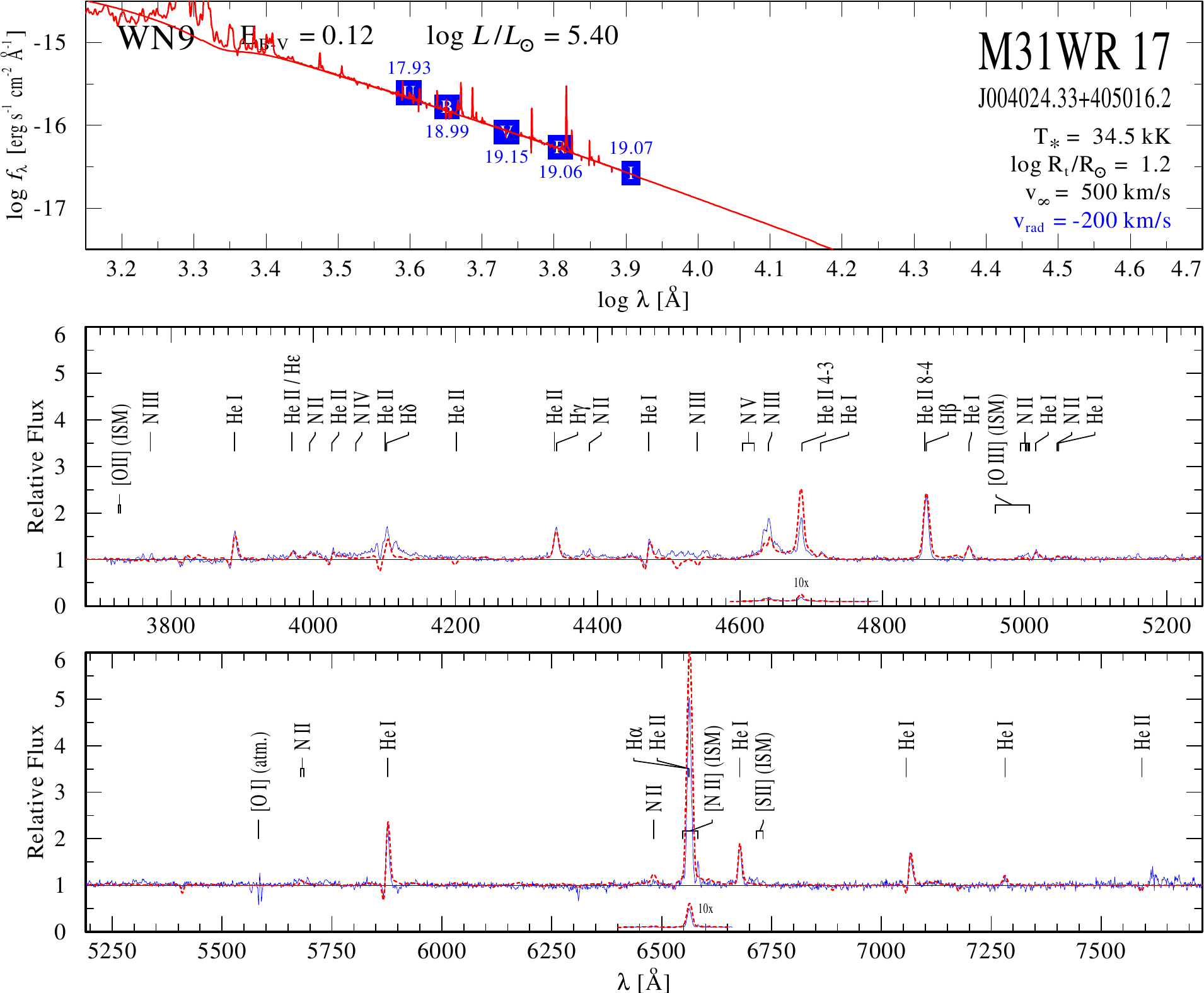}
  \caption{Spectral fit for M31WR\,17}\label{fig:m31wr17}
\end{figure*}

\begin{figure*}
  \centering
  \includegraphics[width=14cm]{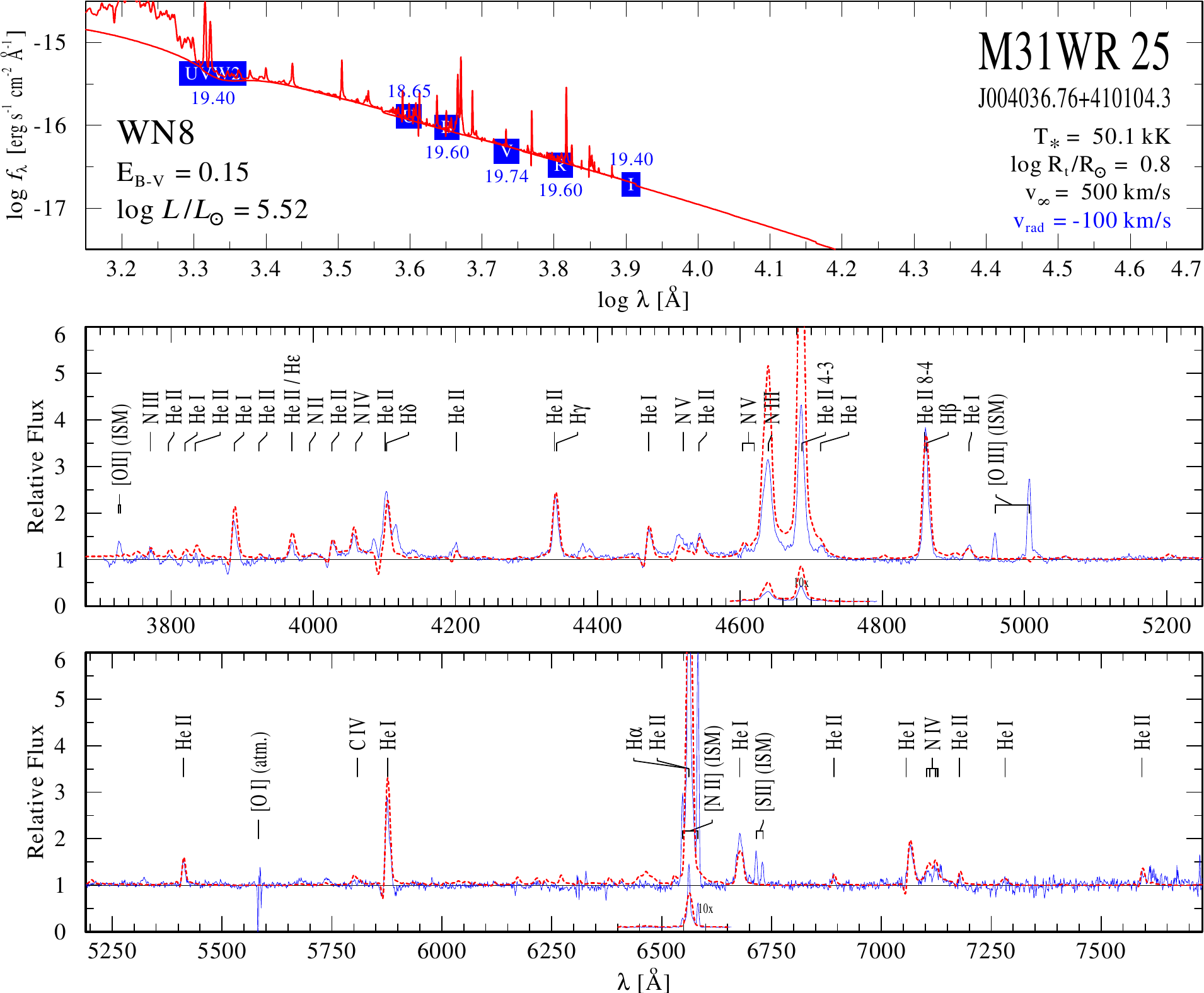}
  \caption{Spectral fit for M31WR\,25}\label{fig:m31wr25}
\end{figure*}

\clearpage

\begin{figure*}
  \centering
  \includegraphics[width=14cm]{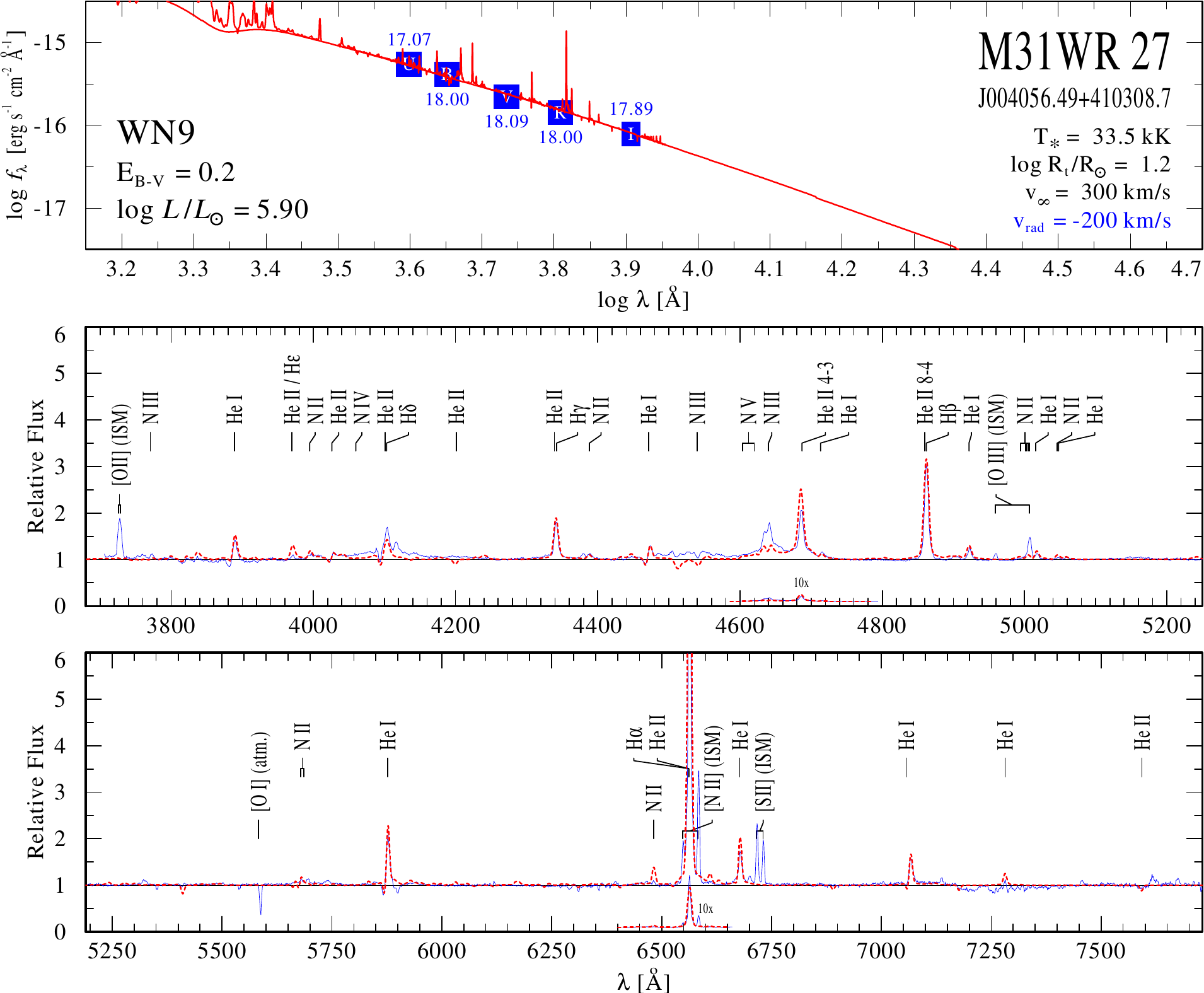}
  \caption{Spectral fit for M31WR\,27}\label{fig:m31wr27}
\end{figure*}

\begin{figure*}
  \centering
  \includegraphics[width=14cm]{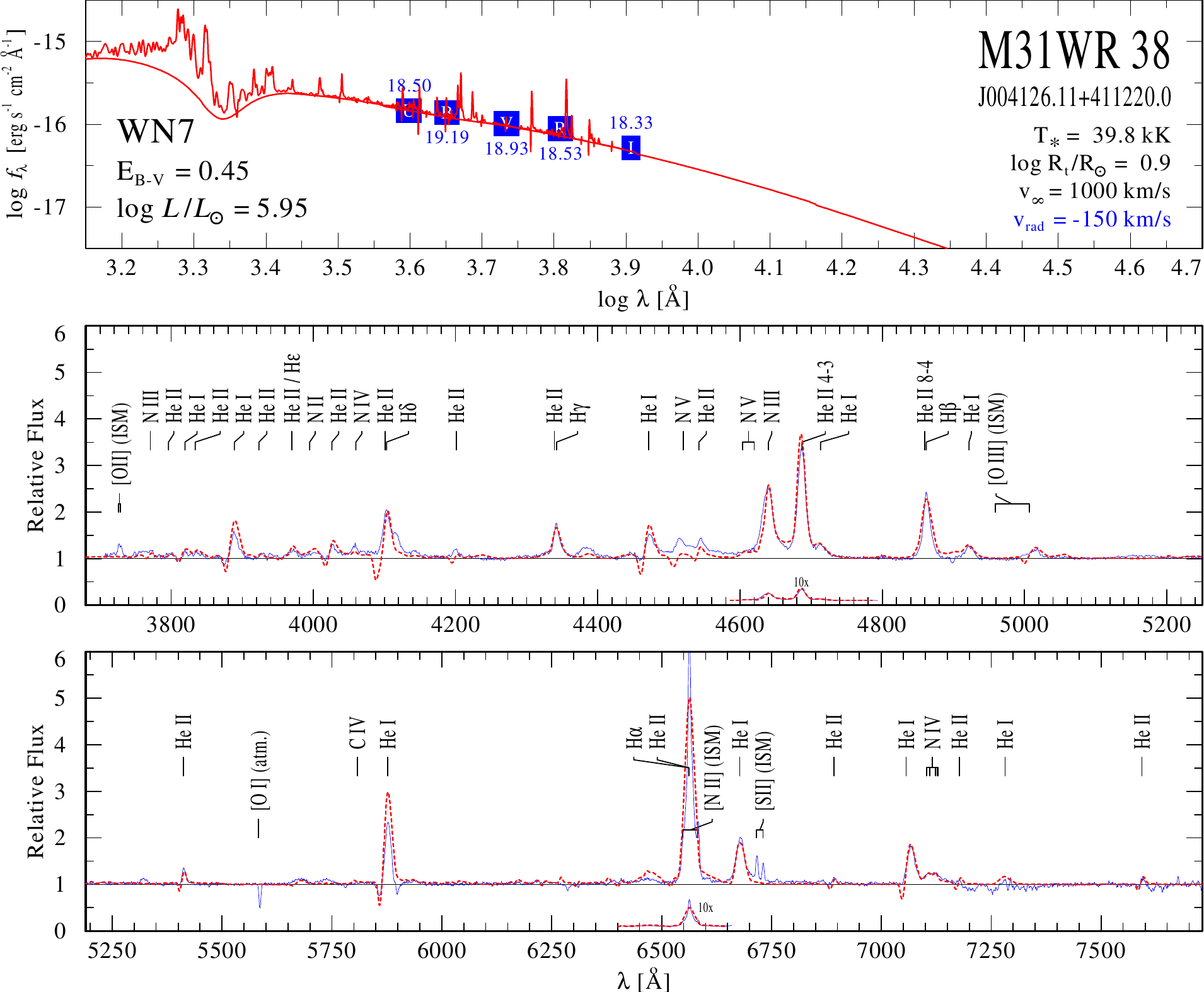}
  \caption{Spectral fit for M31WR\,38}\label{fig:m31wr38}
\end{figure*}

\begin{figure*}
  \centering
  \includegraphics[width=14cm]{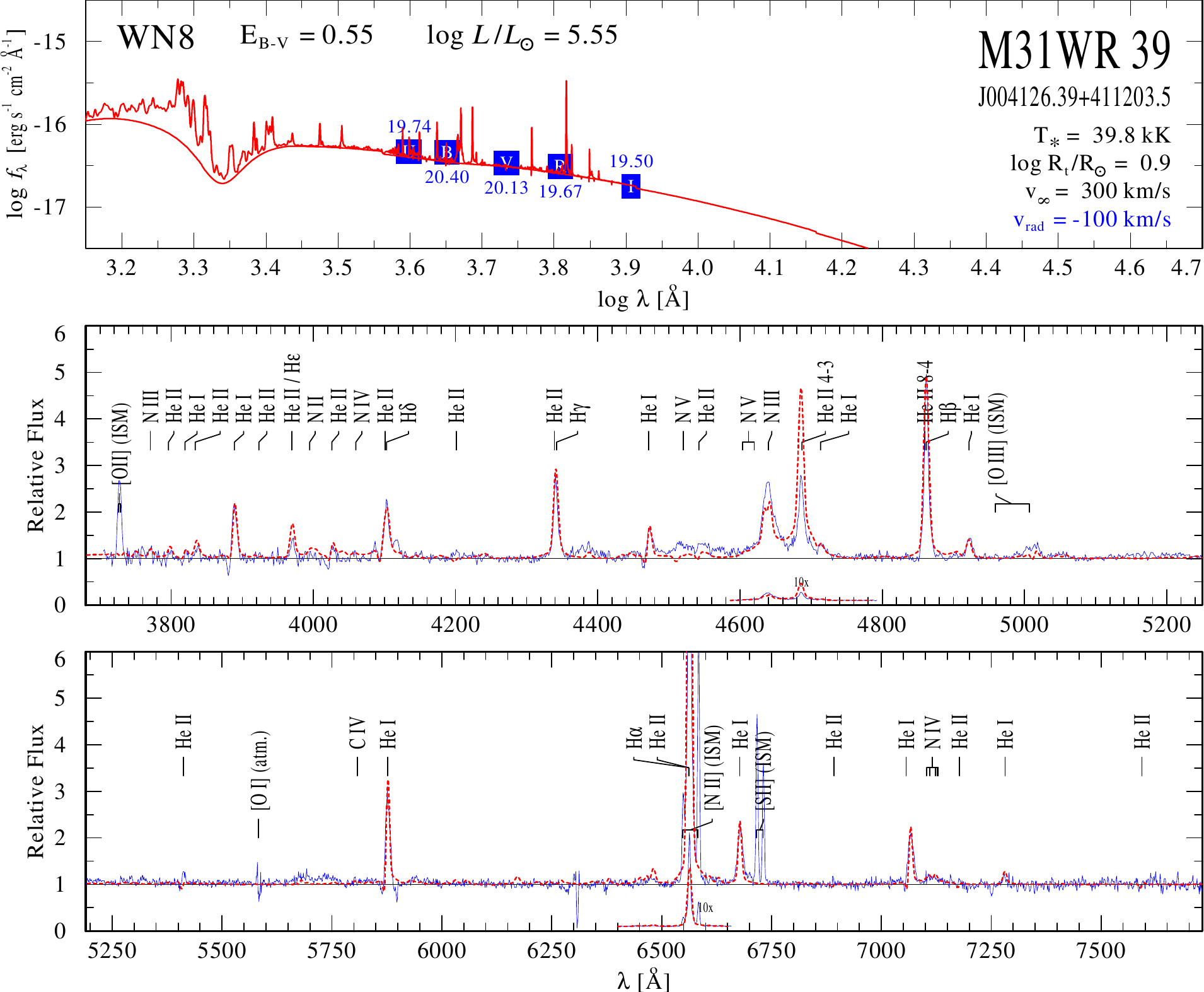}
  \caption{Spectral fit for M31WR\,39}\label{fig:m31wr39}
\end{figure*}

\begin{figure*}
  \centering
  \includegraphics[width=14cm]{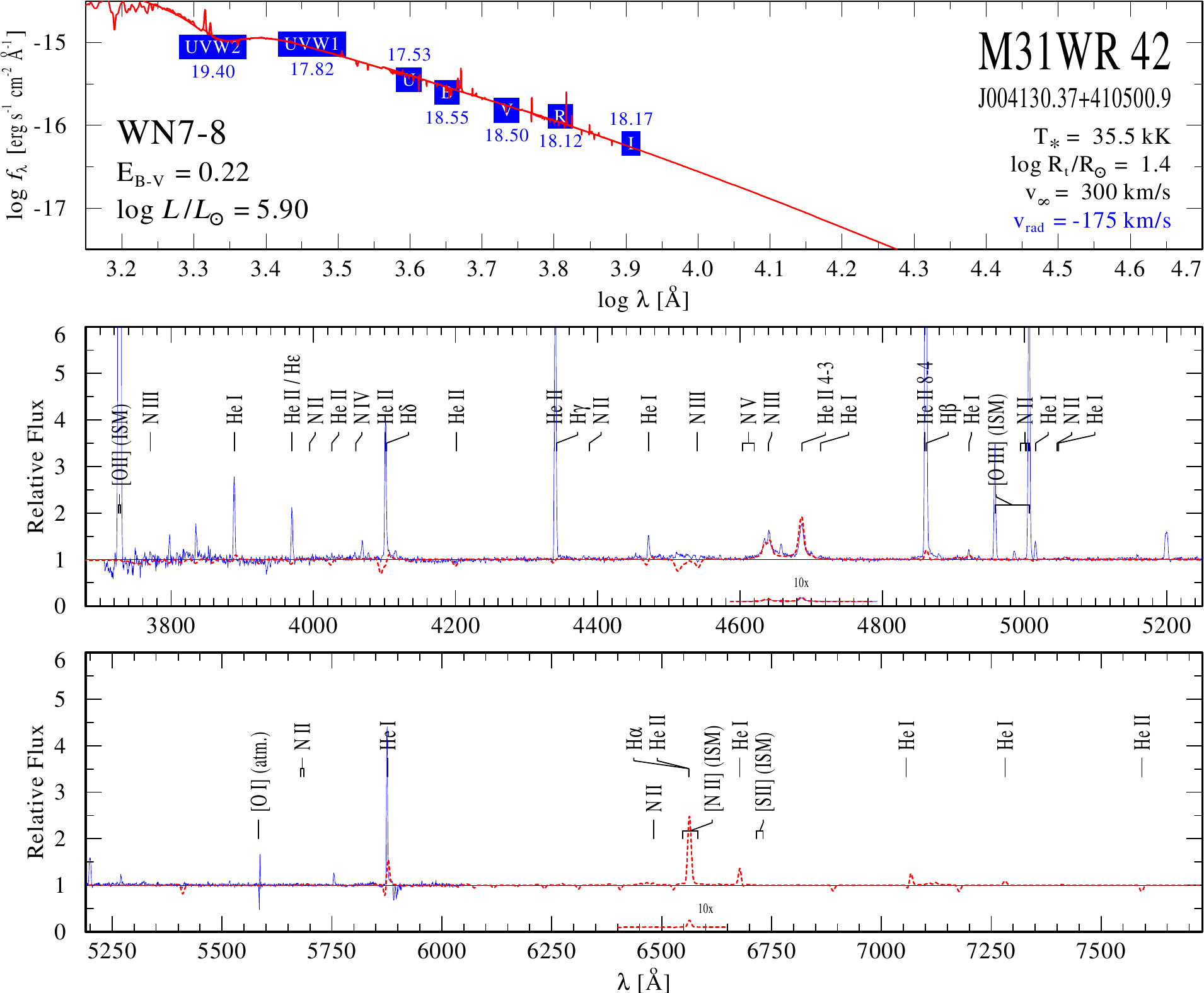}
  \caption{Spectral fit for M31WR\,42}\label{fig:m31wr42}
\end{figure*}

\begin{figure*}
  \centering
  \includegraphics[width=14cm]{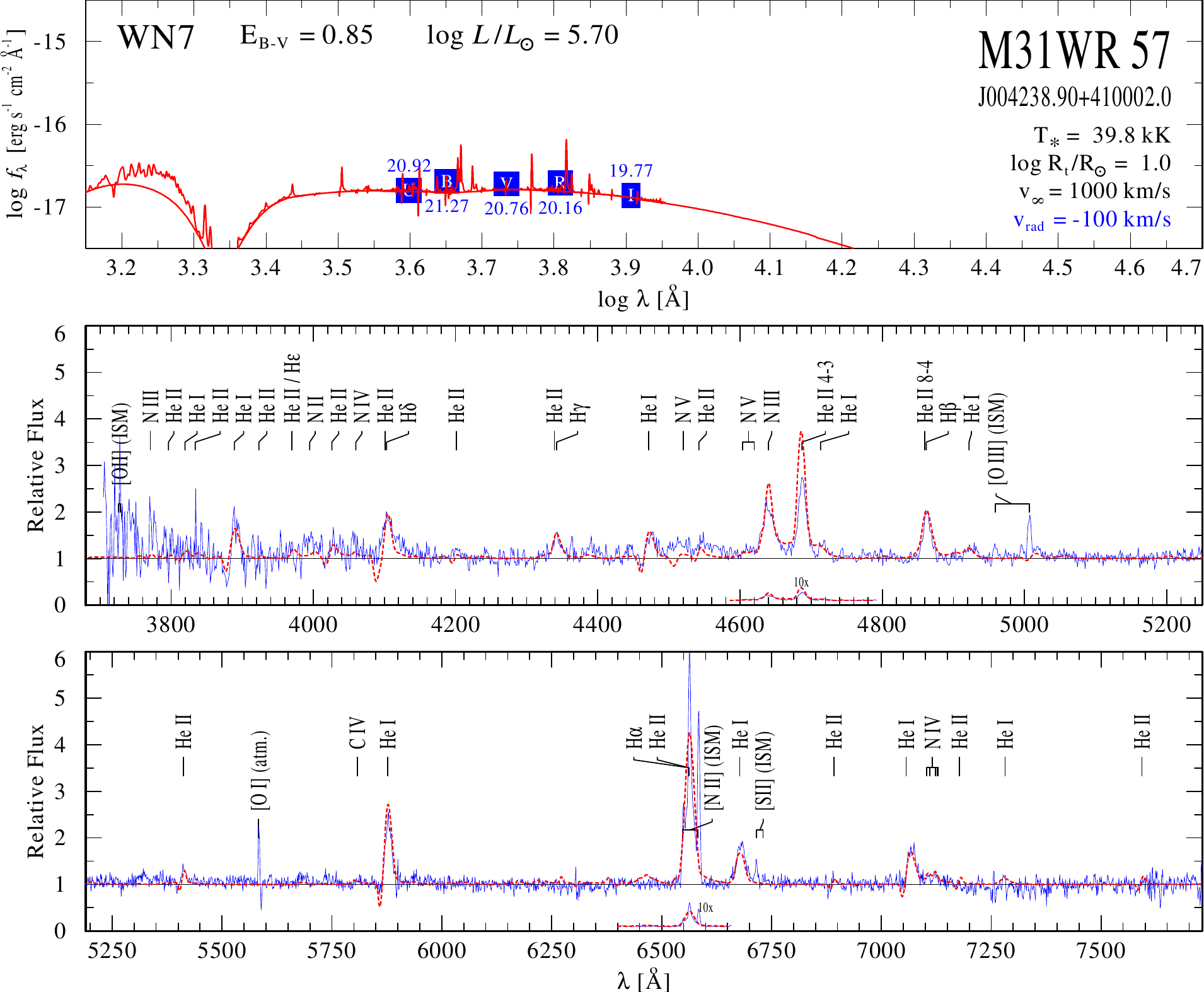}
  \caption{Spectral fit for M31WR\,57}\label{fig:m31wr57}
\end{figure*}

\begin{figure*}
  \centering
  \includegraphics[width=14cm]{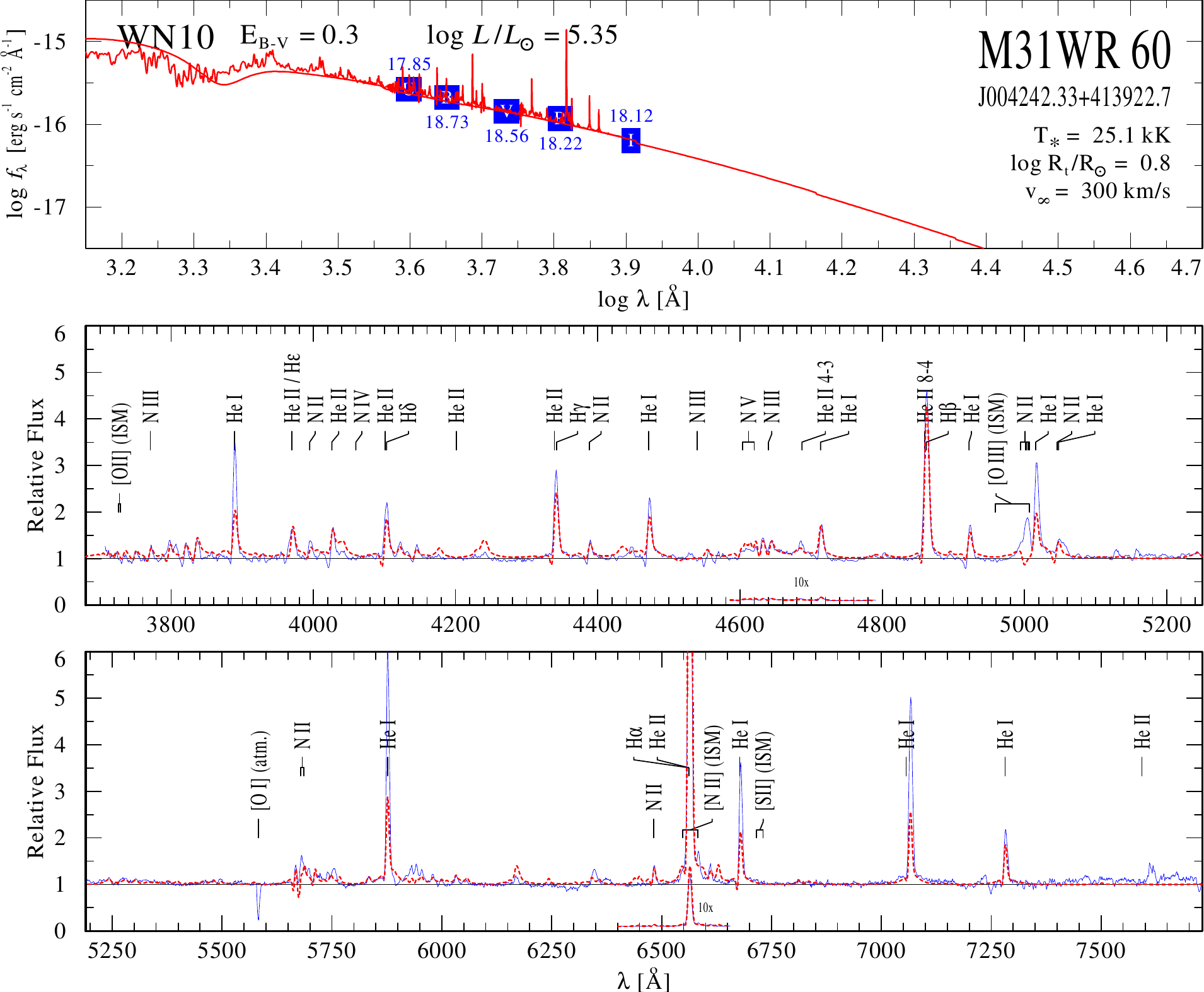}
  \caption{Spectral fit for M31WR\,60}\label{fig:m31wr60}
\end{figure*}

\begin{figure*}
  \centering
  \includegraphics[width=14cm]{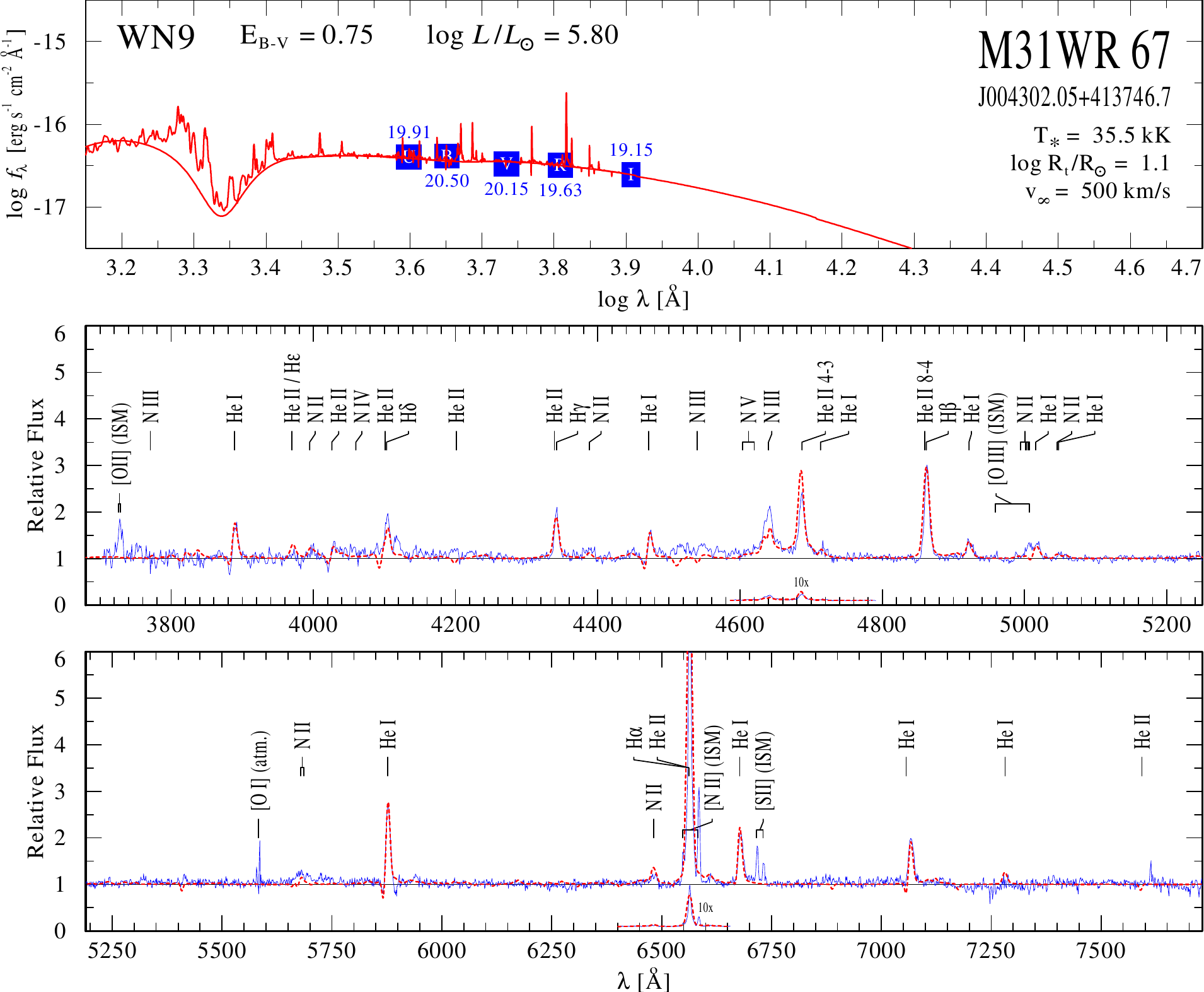}
  \caption{Spectral fit for M31WR\,67}\label{fig:m31wr67}
\end{figure*}

\begin{figure*}
  \centering
  \includegraphics[width=14cm]{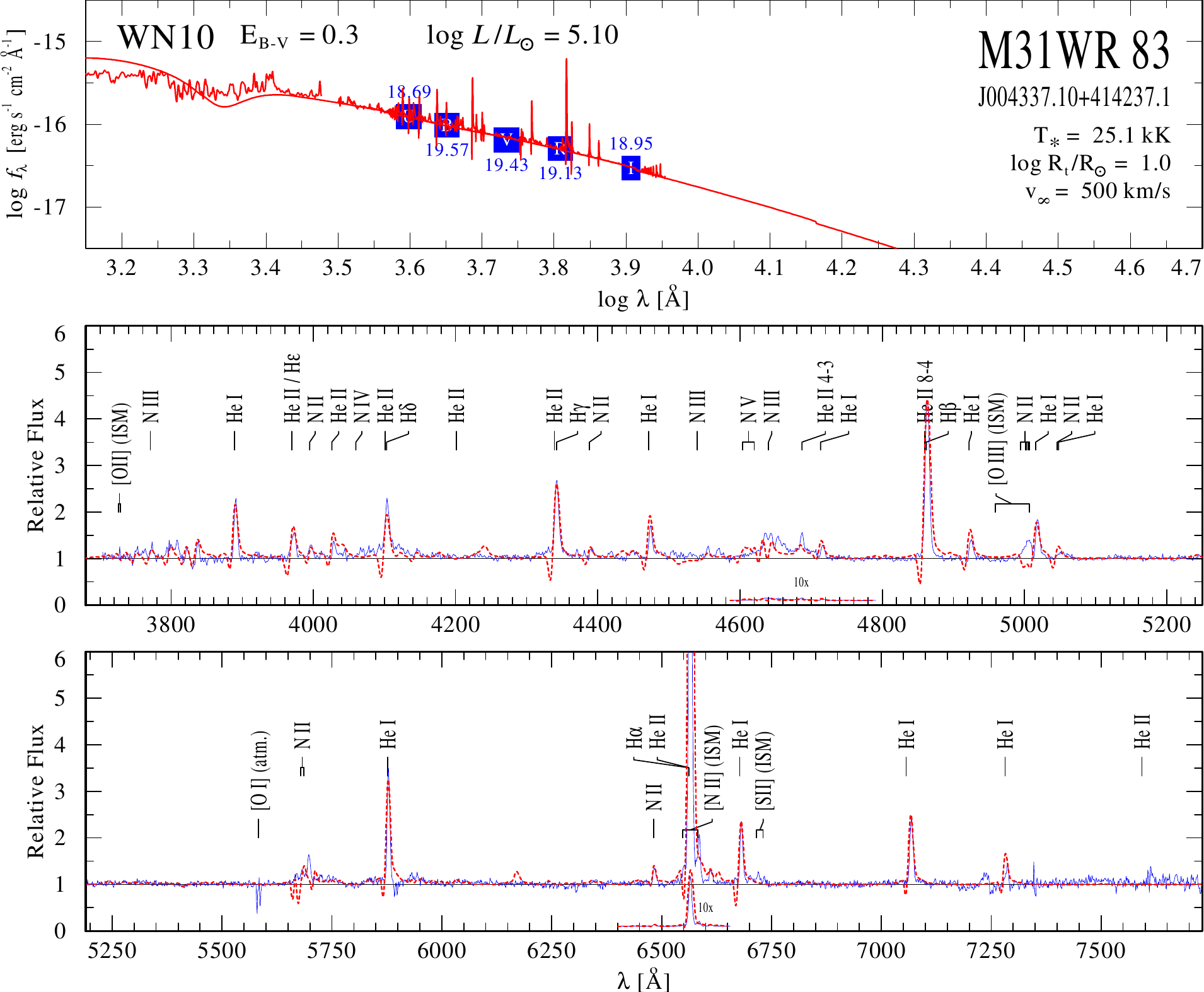}
  \caption{Spectral fit for M31WR\,83}\label{fig:m31wr83}
\end{figure*}

\begin{figure*}
  \centering
  \includegraphics[width=14cm]{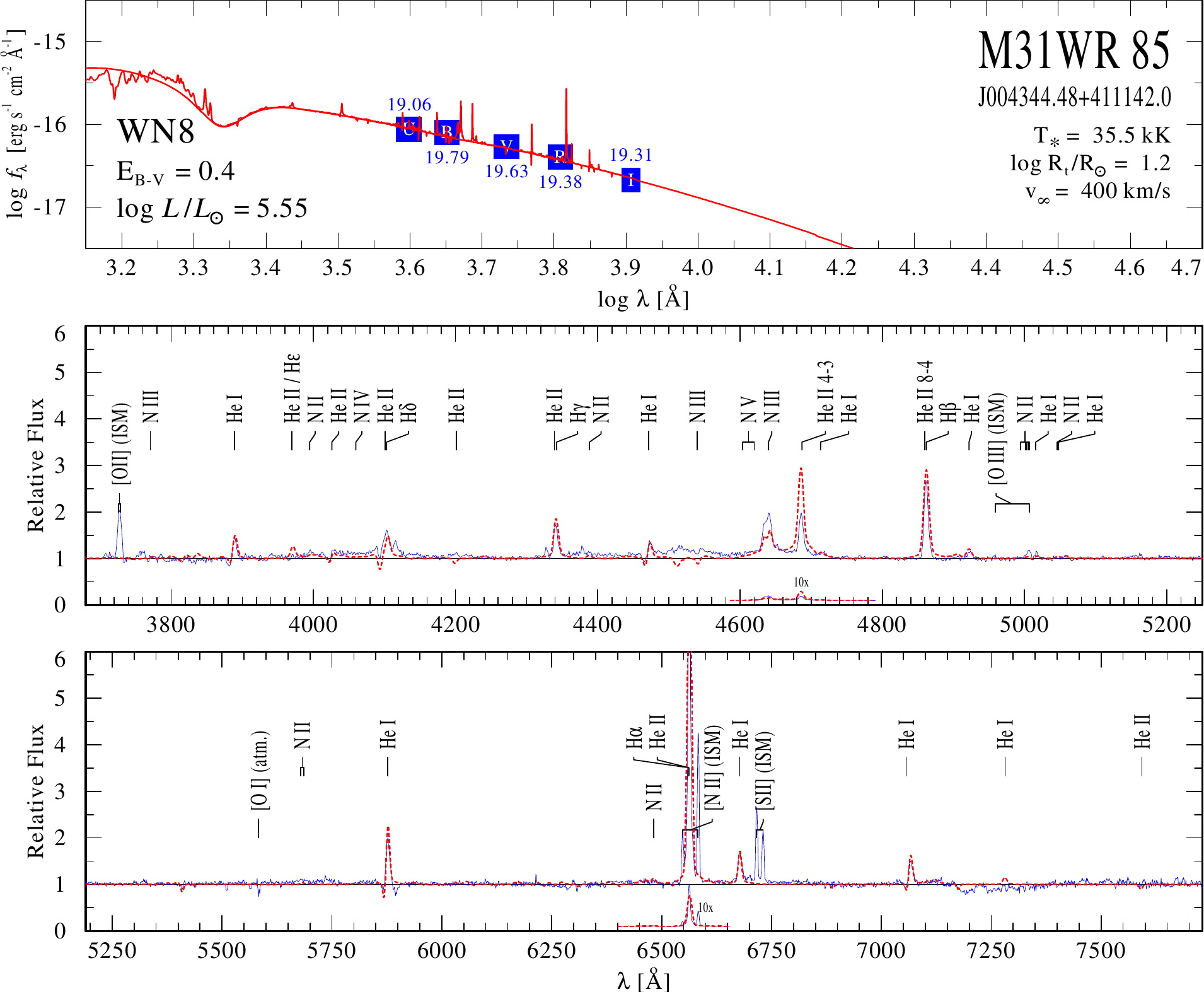}
  \caption{Spectral fit for M31WR\,85}\label{fig:m31wr85}
\end{figure*}

\begin{figure*}
  \centering
  \includegraphics[width=14cm]{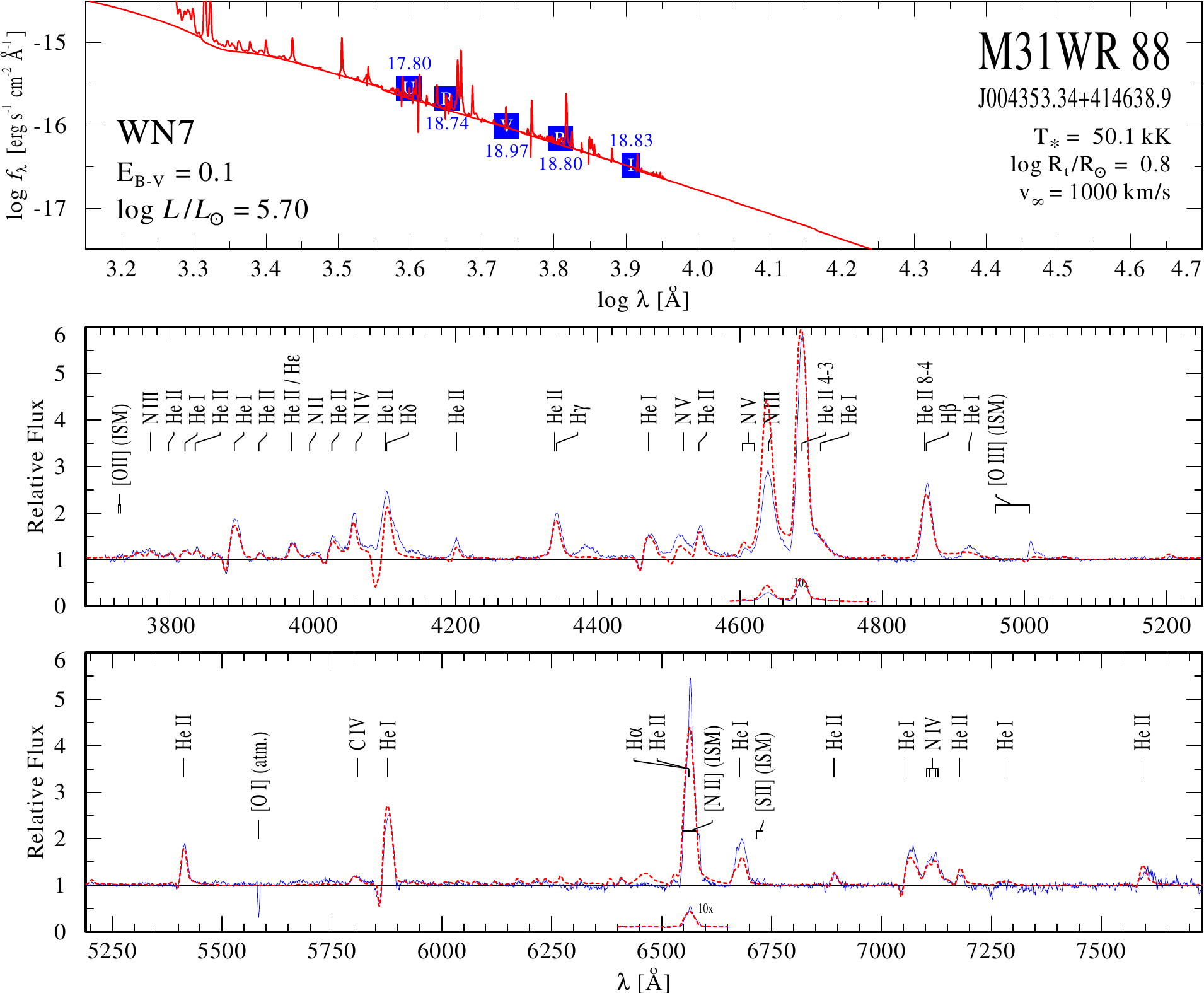}
  \caption{Spectral fit for M31WR\,88}\label{fig:m31wr88}
\end{figure*}

\begin{figure*}
  \centering
  \includegraphics[width=14cm]{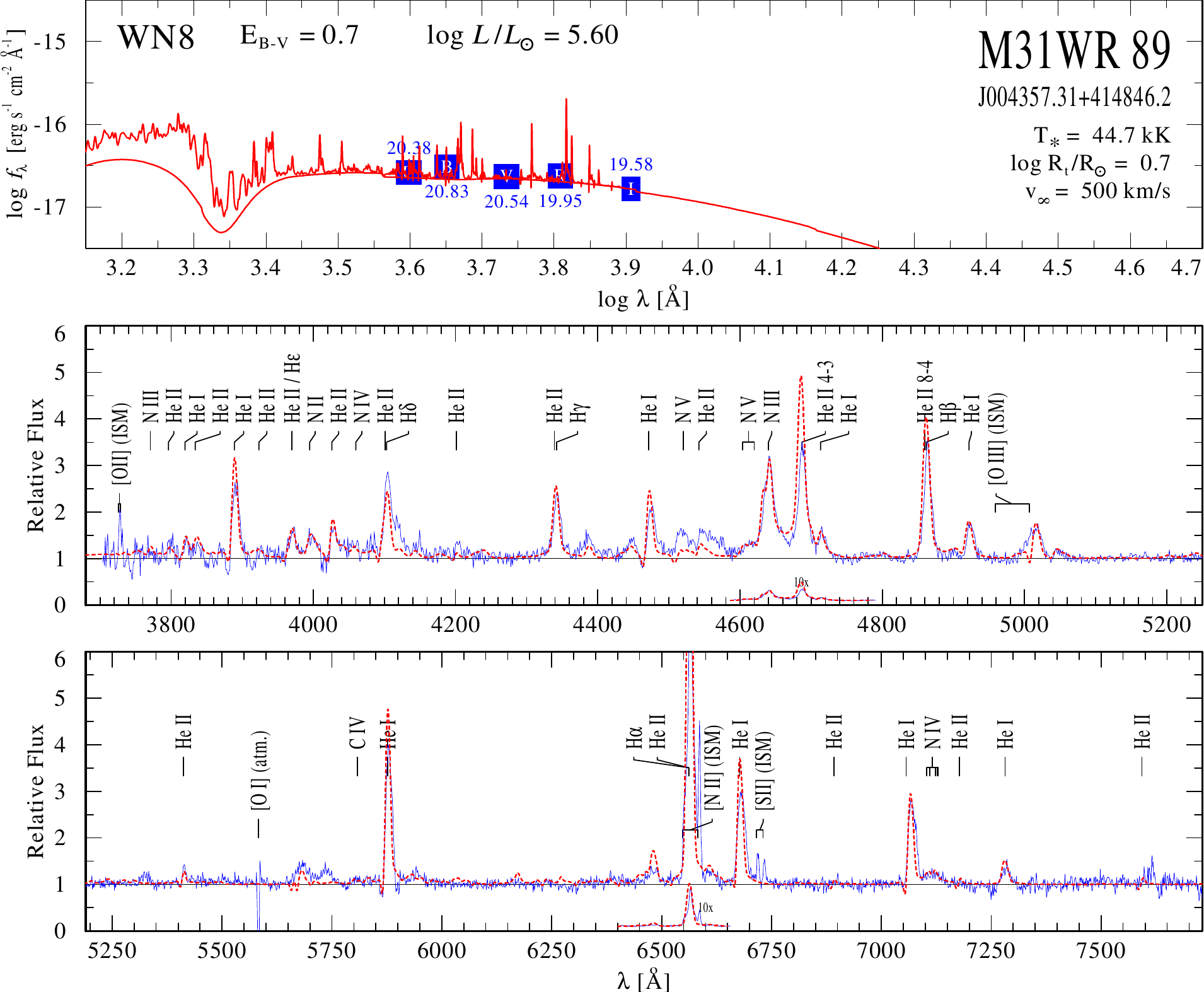}
  \caption{Spectral fit for M31WR\,89}\label{fig:m31wr89}
\end{figure*}

\begin{figure*}
  \centering
  \includegraphics[width=14cm]{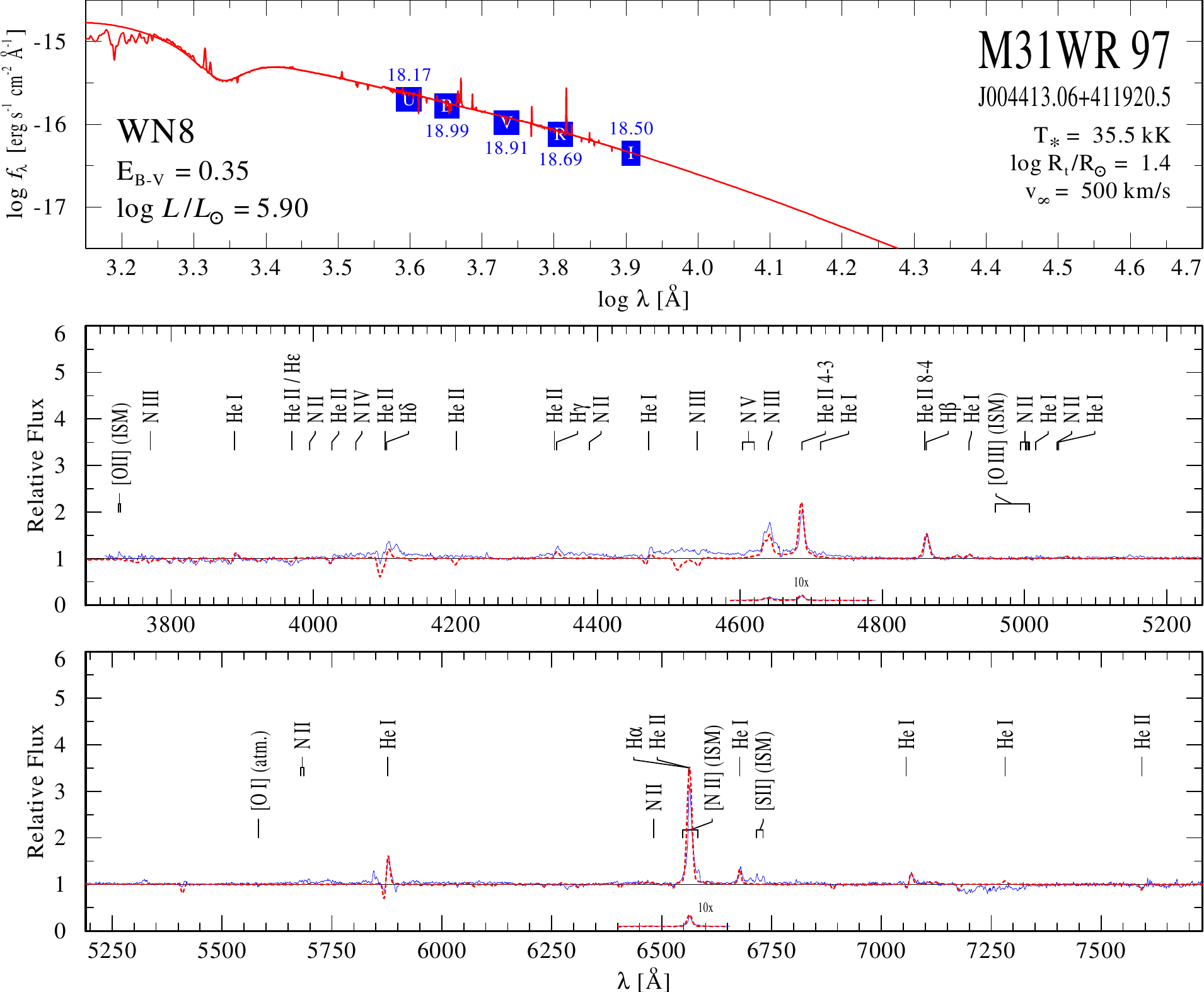}
  \caption{Spectral fit for M31WR\,97}\label{fig:m31wr97}
\end{figure*}

\begin{figure*}
  \centering
  \includegraphics[width=14cm]{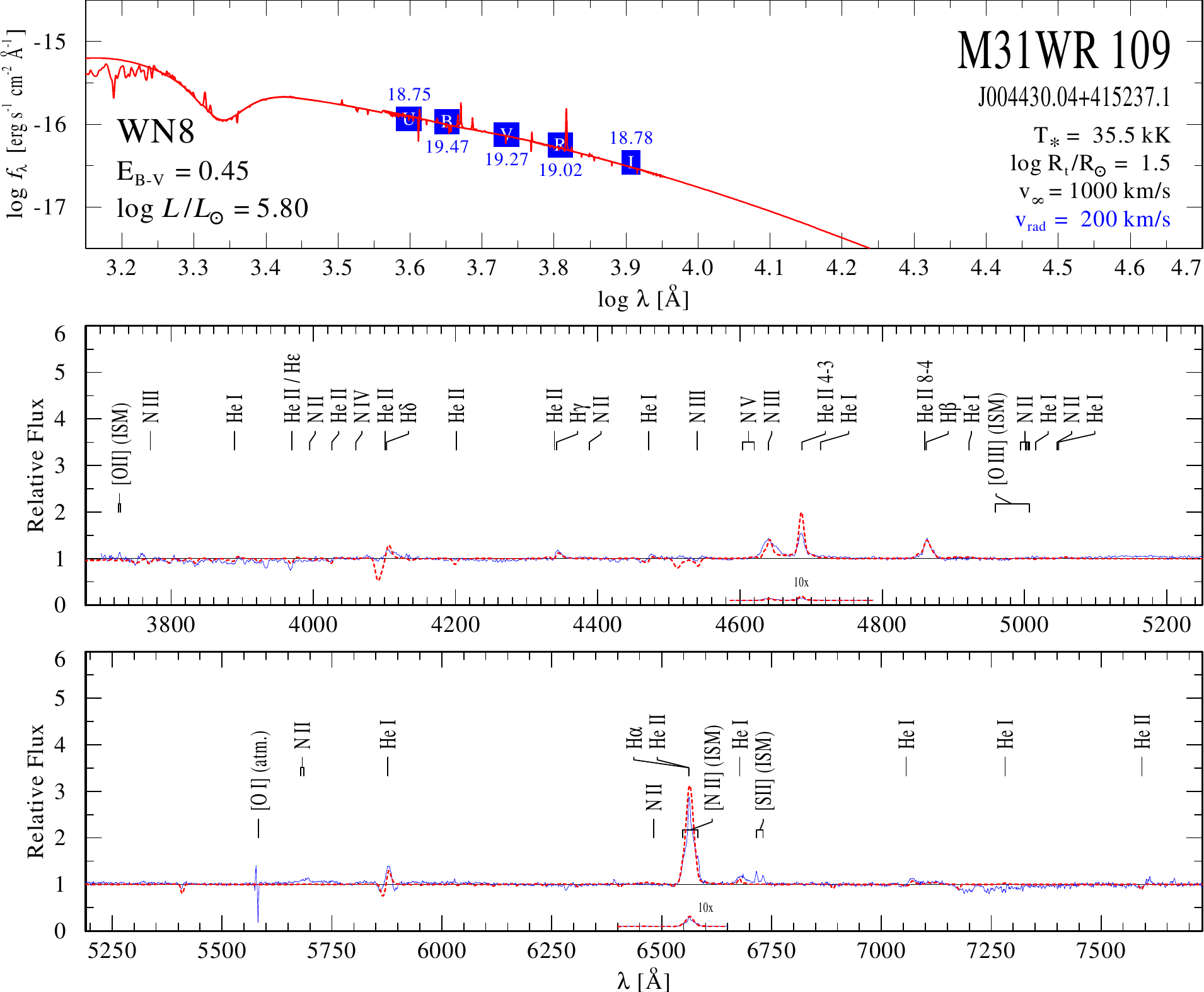}
  \caption{Spectral fit for M31WR\,109}\label{fig:m31wr109}
\end{figure*}

\begin{figure*}
  \centering
  \includegraphics[width=14cm]{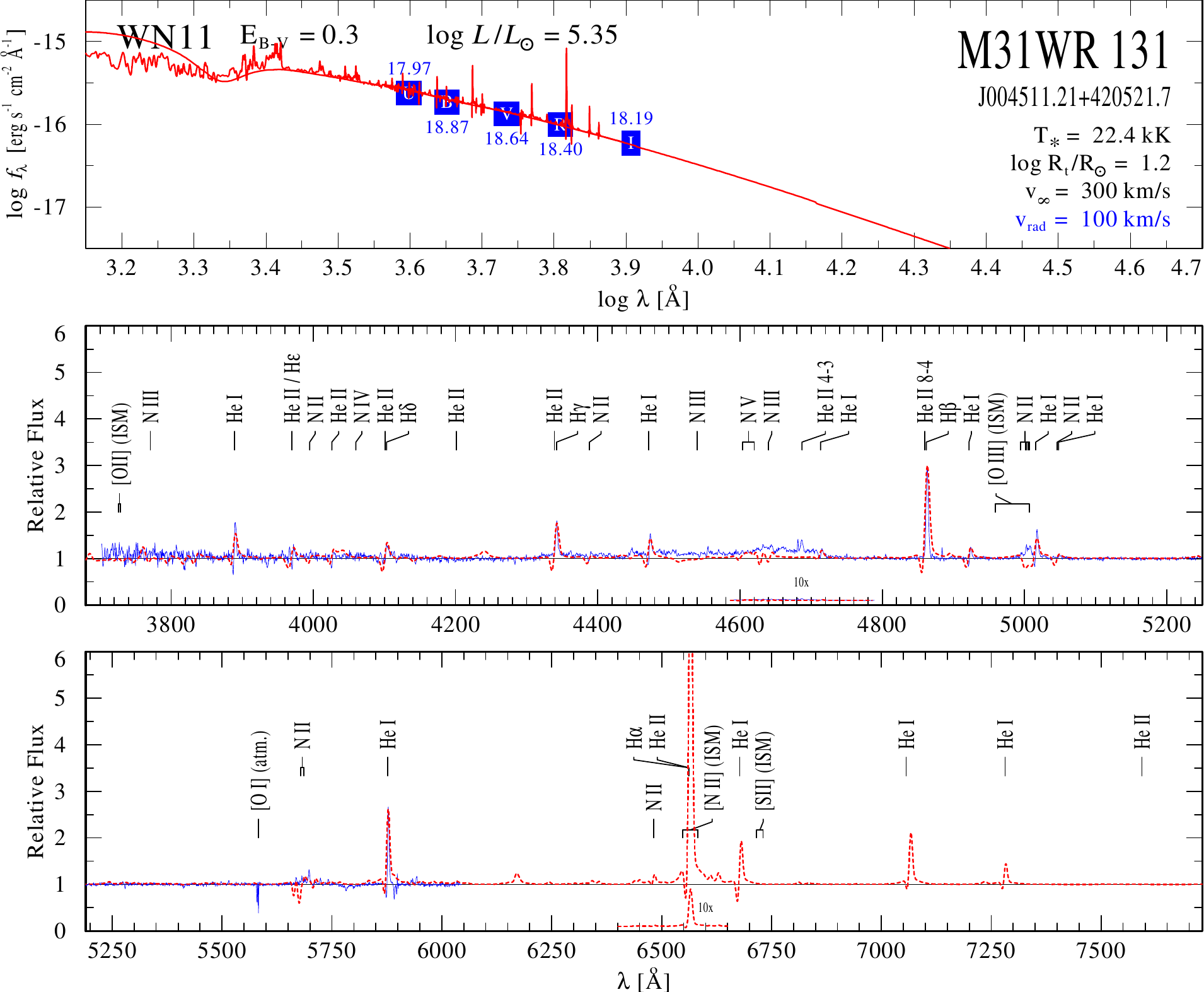}
  \caption{Spectral fit for M31WR\,131}\label{fig:m31wr131}
\end{figure*}

\begin{figure*}
  \centering
  \includegraphics[width=14cm]{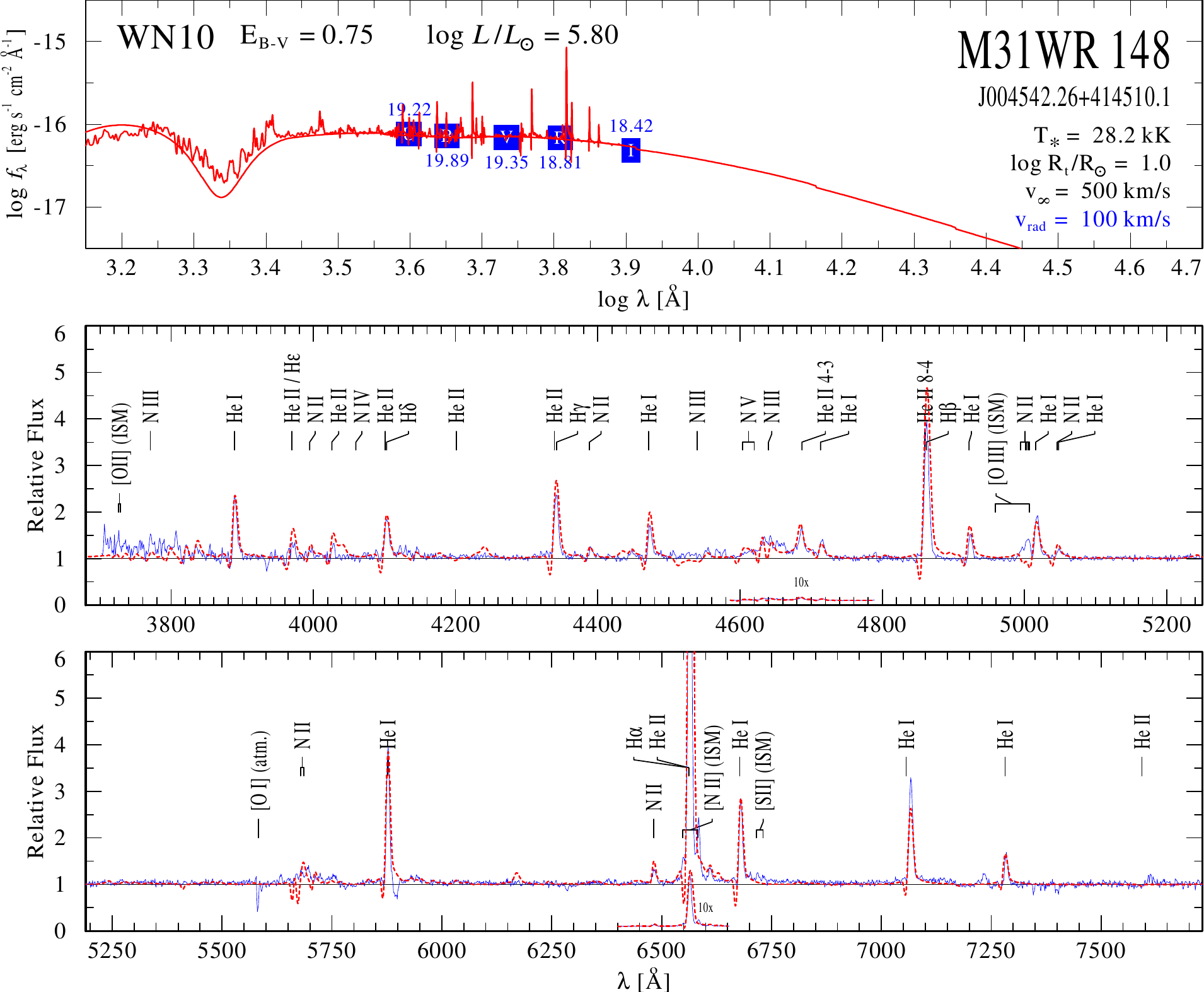}
  \caption{Spectral fit for M31WR\,148}\label{fig:m31wr148}
\end{figure*}

\clearpage

\section{Problematic objects}
  \label{appsec:objnotes}

\subsection{M31WR\,60}

\begin{figure}[htb]
  \resizebox{\hsize}{!}{\includegraphics[angle=0]{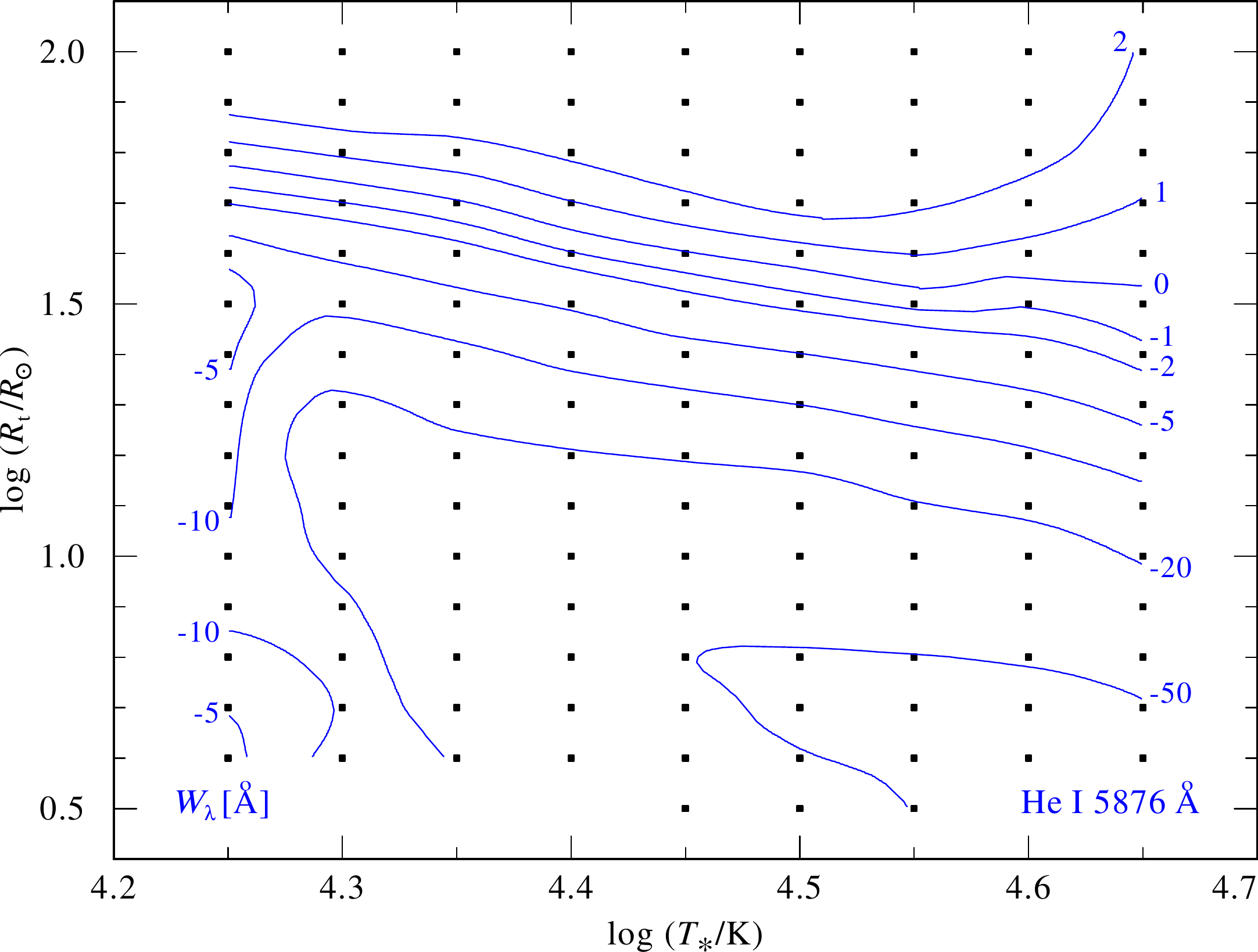}}
  \caption{Contours of constant equivalent width $W_{\lambda}$ for \ion{He}{i} 5876\,\AA\ 
           in the WNh model grid with $X_{\mathrm{H}} = 0.2$ and $\varv_{\infty} = 500\,$km/s.
           A positive value indicates an absorption line, negative values refer
           to emission lines.}
  \label{fig:iso-heI-5876}
\end{figure}

	For the WN10 star \#60 (J004242.33+413922.7), we could not reproduce the peak heights of 
	the unusually strong \ion{He}{i} lines. Looking into the model grids reveals 
	that there seems to be a maximum for the \ion{He}{i} emission lines that 
	cannot be superseded for a given temperature (and chemical composition), even 
	if we further increase the mass-loss rate. This 
	is illustrated in Fig.\,\ref{fig:iso-heI-5876} where we plot contours of 
	constant equivalent width $W_{\lambda}$ for one of our WNh grids. 
	The temperature $T_{\ast}$ is fixed by \ion{He}{ii}\,4686\,\AA\ and \ion{N}{iii}\,4634/42\,\AA, 
	which are reproduced well in our fit (cf. Fig.\,\ref{fig:m31wr60}).
	
	For \#60, we therefore focused on the smaller emission lines that are 
	mostly reproduced. The best result is obtained for a model with 20\%
	hydrogen, even though the H$\beta$, H$\gamma$, and H$\delta$ lines are 
	slightly too weak in the model. However, models with 35\% hydrogen or more
	do not lead to a better fit of the whole spectrum.
	
%--------- Figure   ----------------------------------------------------
%--------- end Figure ----------------------------------------------------
	
\subsection{M31WR\,42}
	
	Another problematic object is \#42, alias J004130.37+410500.9. This spectrum
	is highly contaminated with strong nebular emission. \citet{Massey+2007} refrained
	from assigning a subtypes and listed the star simply as WNL. \citet{NMG2012} speculated 
	that it might be of WN7 subtype, but have not included this in their table 
	either. Unfortunately
	the observed spectrum ends at $6000\,$\AA\ so that H$\alpha$ is not covered.
	 Only two prominent WN features are available for the diagnostic, 
	\ion{N}{iii}\,4634/42\,\AA\ and \ion{He}{ii}\,4686\,\AA, both in emission. 
	Indeed the roughly similar peak heights of these two lines would 
	classify the star as WN7 following the scheme of \citet{vdH2001}, but since 
	\ion{N}{iv} is so weak in the spectrum, it could also be classified as WN8. 
	However, we do not see strong P-Cyg-profiles that would be 
	another criterion for the WN8 subtype. We therefore specify the 
	subtype as WN7-8 until future observations become available. 
	
	From the two diagnostic lines we can obtain the basic parameters $T_{\ast}$ 
	and $R_{\mathrm{t}}$ for \#42. Owing to the strong nebular 
	emission lines, the hydrogen content $X_{\mathrm{H}}$ and the terminal 
	velocity $\varv_{\infty}$ are harder to determine. From the unblended lines,
	neither the \ion{N}{iii}-complex at 4634-42\,\AA\ nor the weak \ion{He}{ii}\,4686\,\AA\ line 
	are really sensitive to $\varv_{\infty}$. We therefore use the \ion{H}{i} and 
	\ion{He}{i} lines, although they are polluted with nebular emission. 
	We can rule out velocities higher than $\varv_{\infty} = 500\,$km/s 
	because they would produce line profiles that are broader than the observed lines. 
	The best compromise is obtained with $\varv_{\infty} = 300\,$km/s, which would 
	favor WN8 over the WN7 subtype classification, where we usually find 
	higher terminal velocities. 
	
	For the hydrogen content, we use the indirect sensitivity of \ion{He}{ii}\,4686\,\AA\ 
	to rule out models with	50\% hydrogen and zero hydrogen. We favor a model with 20\% 
	over those with 35\% hydrogen because we observe a small emission of \ion{C}{iii}\,4650\,\AA\ and 
	therefore expect a more ``evolved'' star, but a larger uncertainty than for 
	the other objects remains.

\end{appendix}

\end{document}